\documentclass[a4paper, 9pt]{article}

\usepackage[american]{babel}
\usepackage{epsfig}
\usepackage{amssymb}
\usepackage{array}
\usepackage{multicol}
\usepackage{vmargin}

\setpapersize[portrait]{A4}
\setmarginsrb{1.35cm}{0.3cm}{1.05cm}{0.8cm}{0.9cm}{0.9cm}{0.9cm}{0.9cm}

\newcommand{\angstrom}{\textnormal{\AA}}

\newcommand{\area}{\ensuremath{\mathnormal{A}}}

\newcommand{\absnum}{\ensuremath{\mathnormal{N}}}
\newcommand{\boost}{\textbf{\textsf{boost}}}

\newcommand{\chempot}{\ensuremath{\mathnormal{\mu}}}
\newcommand{\chempotsat}{\ensuremath{\chempot_\mathnormal{\sigma}}}

\newcommand{\cv}{\ensuremath{\mathnormal{c}_\mathnormal{v}}}
\newcommand{\critnuclarea}{\ensuremath{\mathnormal{\area^*}}}
\newcommand{\critnucldensity}{\ensuremath{\mathnormal{\numdensity^*}}}
\newcommand{\critsize}{\ensuremath{\mathnormal{\iota^*}}}

\newcommand{\csizea}{\ensuremath{\mathnormal{\iota}}}
\newcommand{\csizeb}{\ensuremath{\textnormal{n}}}
\newcommand{\DeltaTcrit}{\ensuremath{\mathnormal{\Delta\temperature^*}}}
\newcommand{\distance}{\ensuremath{\mathnormal{r}}}
\newcommand{\distancecut}{\ensuremath{\distance_\textnormal{\tiny{}c}}}
\newcommand{\elongation}{\ensuremath{\mathnormal{L}}}
\newcommand{\energydeviation}{\ensuremath{\mathnormal{b}}}
\newcommand{\enthalpy}{\ensuremath{\mathnormal{\Delta{}H_v}}}
\newcommand{\erf}{\ensuremath{\textnormal{erf}}}
\newcommand{\frontier}{\ensuremath{\mathnormal{\nu}}}

\newcommand{\gibbsenergy}{\ensuremath{\mathnormal{G}}}
\newcommand{\impactrate}{\ensuremath{\mathnormal{\beta}}}
\newcommand{\inta}{\ensuremath{\mathnormal{i}}}
\newcommand{\intb}{\ensuremath{\mathnormal{j}}}

\newcommand{\internal}{\ensuremath{\mathnormal{u}}}
\newcommand{\kboltz}{\ensuremath{\mathnormal{k}}}

\newcommand{\lfkalpha}{\ensuremath{\mathnormal{\alpha}}}
\newcommand{\lfkK}{\ensuremath{\mathnormal{\kappa}}}
\newcommand{\lfktau}{\ensuremath{\mathnormal{\tau}}}
\newcommand{\lfkTheta}{\ensuremath{\mathnormal{\Theta}}}
\newcommand{\lfkqzero}{\ensuremath{\mathnormal{q}_0}}
\newcommand{\liqnumdensity}{\ensuremath{\numdensity_\textnormal{\tiny{}l}}}

\newcommand{\LJepsilon}{\ensuremath{\mathnormal{\epsilon}}}
\newcommand{\LJsigma}{\ensuremath{\mathnormal{\sigma}}}
\newcommand{\lsone}{\textbf{\textsf{ls$_1$}}}

\newcommand{\mfpt}{\ensuremath{\mathnormal{\digamma}}}

\newcommand{\molecularmass}{\ensuremath{\mathnormal{m}}}
\newcommand{\naturals}{\ensuremath{\mathbb{N}}}
\newcommand{\ndsat}{\ensuremath{\numdensity_\mathnormal{\sigma}}}
\newcommand{\nuclrate}{\ensuremath{\mathnormal{J}}}
\newcommand{\noniso}{\ensuremath{\mathnormal{\vartheta}}}
\newcommand{\numdensity}{\ensuremath{\mathnormal{\rho}}}

\newcommand{\planartension}{\ensuremath{\surfacetension_0}}
\newcommand{\potential}{\ensuremath{\mathnormal{u}}}
\newcommand{\potentialLJ}{\ensuremath{\potential_\textnormal{\tiny{}LJ}}}

\newcommand{\pressure}{\ensuremath{\mathnormal{p}}}
\newcommand{\psat}{\ensuremath{\pressure_\mathnormal{\sigma}}}
\newcommand{\quadrupole}{\ensuremath{\mathnormal{Q}}}
\newcommand{\clusterradius}{\mathnormal{r}}
\newcommand{\reala}{\ensuremath{\mathnormal{x}}}
\newcommand{\realb}{\ensuremath{\mathnormal{\xi}}}
\newcommand{\releaseenergy}{\ensuremath{\mathnormal{q}}}
\newcommand{\satnumdensity}{\ensuremath{\numdensity_{\tiny{}\sigma}}}
\newcommand{\specenergy}{\ensuremath{\mathcal{E}}}

\newcommand{\stillingerdistance}{\ensuremath{\mathnormal{r}_\textnormal{\small{}gc}}}
\newcommand{\super}{\ensuremath{\mathcal{S}}}
\newcommand{\supermu}{\ensuremath{\super_\chempot}}
\newcommand{\superp}{\ensuremath{\super_\pressure}}
\newcommand{\superrho}{\ensuremath{\super_\numdensity}}
\newcommand{\supersat}{\superp}
\newcommand{\surfaceenergy}{\ensuremath{\mathnormal{\phi}}}
\newcommand{\surfacetension}{\ensuremath{\mathnormal{\gamma}}}
\newcommand{\temperature}{\ensuremath{\mathnormal{T}}}

\newcommand{\timea}{\ensuremath{\mathnormal{t}}}

\newcommand{\vapnumdensity}{\ensuremath{\numdensity_\textnormal{\tiny{}v}}}

\newcommand{\velocity}{\ensuremath{\mathnormal{v}}}
\newcommand{\virial}{\ensuremath{\mathcal{B}}}
\newcommand{\volume}{\ensuremath{\mathnormal{V}}}

\newcommand{\zeldovich}{\ensuremath{\mathcal{Z}}}

\newcommand{\qq}[1]{{\lq}#1{\rq}}

\title{Homogeneous nucleation in supersaturated vapors of methane, ethane, and
       carbon dioxide predicted by brute force molecular dynamics}
\date{}
\author{Martin Horsch$^\dagger$,
   Jadran Vrabec\footnote{Corresponding author. E-mail: vrabec@itt.uni-stuttgart.de}
      \footnote{Universit\"at Stuttgart, Institute of Thermodynamics and Thermal
         Process Engineering, Pfaffenwaldring 9, 70569 Stuttgart, Germany}~,
   Martin Bernreuther\footnote{High Performance Computing Center Stuttgart,
      Nobelstr.\ 19, 70550 Stuttgart, Germany}~,
   Sebastian Grottel\footnote{Universit\"at Stuttgart, Institute of Visualization and Interactive Systems,
      Universit\"atsstr.\ 38, 70569 Stuttgart, Germany}~, \\
   Guido Reina$^\mathsection$, 
   Andrea Wix\footnote{Universit\"at Karls\-ruhe, Institut f\"ur Technische Thermodynamik
      und K\"altetechnik, Engler-Bunte-Ring 21, 76131 Karlsruhe, Germany}~,
   Karlheinz Schaber$^\mathparagraph$,
   and Hans Hasse$^\dagger$}

\begin{document}
\maketitle
\begin{abstract}
\noindent Molecular dynamics (MD) simulation is applied to the condensation process of supersaturated vapors
of methane, ethane, and carbon dioxide.
Simulations of systems with up to a million particles were conducted with a massively parallel MD program.
This leads to reliable statistics and makes nucleation rates down to the order of $10^{30}$ m$^{-3}$s$^{-1}$
accessible to the direct simulation approach.
Simulation results are compared to the classical nucleation theory (CNT) as well
as the theory of Laaksonen, Ford, and Kulmala (LFK) which introduces a size dependence of the specific surface
energy. CNT describes the nucleation of ethane and carbon dioxide excellently over the entire studied
temperature range, whereas LFK provides a better approach to methane at low temperatures.
\end{abstract}

\section{\bf\large Introduction}
\label{sec:introduction}
\begin{multicols}{2}
Homogeneous nucleation was discussed theoretically
by Gibbs \cite{Gibbs28} and studied in depth by Volmer and Weber \cite{VW26}
as well as Farkas \cite{Farkas27}. In combination with experiments carried out
by Wilson \cite{Wilson97} and Powell \cite{Powell28} during the same period, these efforts
established the classical nucleation theory (CNT), which is known to be accurate in many
cases but to fail in others \cite{FRKBU98, Gunton99, Ford04}.

Molecular simulations are applied to this problem since the late 1950s,
when Alder and Wainwright \cite{AW57} observed a first-order phase transition
in molecular dynamics (MD) simulations of the hard sphere fluid.
In the 1970s, McGinty \cite{McGinty73} studied liquid clusters of the
{L}ennard-{J}ones (LJ) fluid in MD simulations, and
Rao et al.\ \cite{RBK78} described the condensation of a supersaturated vapor
with results obtained from both Monte Carlo (MC) and MD simulations.
Some common approaches to the dynamics of nucleation, such as 
MD simulations with
inserted droplets \cite{Zhukovitskii95, LWSL00, HK03, Barrett07, SNV07}
or transition path sampling \cite{AWW05, TDP06, EMB03}
as well as MC simulations \cite{EMB03, VF00, NV05, BCB07, MZLV07},
do not lead immediately to the velocity of the phase transition, but only to {\em indirect}
information, e.g.\ on the required activation energy.
The present study discusses brute force MD simulations, which are aimed at the {\em direct} reproduction
of a nucleation process by means of the deterministic simulation of a large system.

Nucleation processes are characterized by the nucleation rate $\nuclrate$, i.e.\ the number of stable liquid nuclei
generated per volume and time, and their critical size $\critsize$, i.e.\ the number of molecules in a
nucleus with maximal Gibbs energy of formation.
Droplets above that size have a higher probability to further grow, whereas smaller clusters tend to evaporate.
Due to current limitations in the available computational resources, only nucleation processes with
extremely large values of $\nuclrate$ can be simulated by MD.
However, when nucleation occurs very rapidly, the vapor phase is not fully in equilibrium
with the emerging droplets and the critical size is not constant.
It is nonetheless possible to determine nucleation
rates if one follows the somewhat heuristic approach proposed
by Yasuoka and Matsumoto \cite{YM98}. Most recent direct MD studies of
nucleation \cite{KAAA05, TKTN05a, Kraska06, LK07} adhere to this method.

The method of Yasuoka and Matsumoto requires system sizes and simulation times to be as large as possible.
Due to restrictions of computational power, the lowest nucleation rates which  
can be obtained nowadays with this approach -- above $10^{30} \slash (\textnormal{m}^3\textnormal{s})$ in the 
present study -- exceed those which actually can be observed in experiments by
about seven orders of magnitude \cite{Iland04}.
This gap can only be closed by predictions on the basis of nucleation theories that
express the dependence of $\nuclrate$ and $\critsize$ on temperature and pressure, where the latter is often given
in terms of the supersaturation $\supersat(\pressure, \temperature) = \pressure \slash \psat(\temperature)$,
i.e.\ related to the vapor pressure $\psat$.
Reviews following the progress of the last decades are provided by Oxtoby \cite{Oxtoby92, Oxtoby98}
and Ford \cite{Ford04}.
For a description of advanced experimental methods see also Fladerer and Strey \cite{FS06}
as well as Iland \cite{Iland04}.

In the following sections, CNT and a version of the Dillmann-Meier \cite{DM90}
model due to Laaksonen, Ford, and Kulmala \cite{LFK94}, referred to as LFK here,
will be discussed and compared to data from direct MD simulations.
It is also necessary to comment on the {\em mean first passage time} (MFPT) approach, an indirect
method which fits a predefined kinetic model with three parameters to simulation results \cite{BW06, WSR07}.
\end{multicols}

\begin{figure}[h]
\centerline{
   \includegraphics[width=5.67cm]{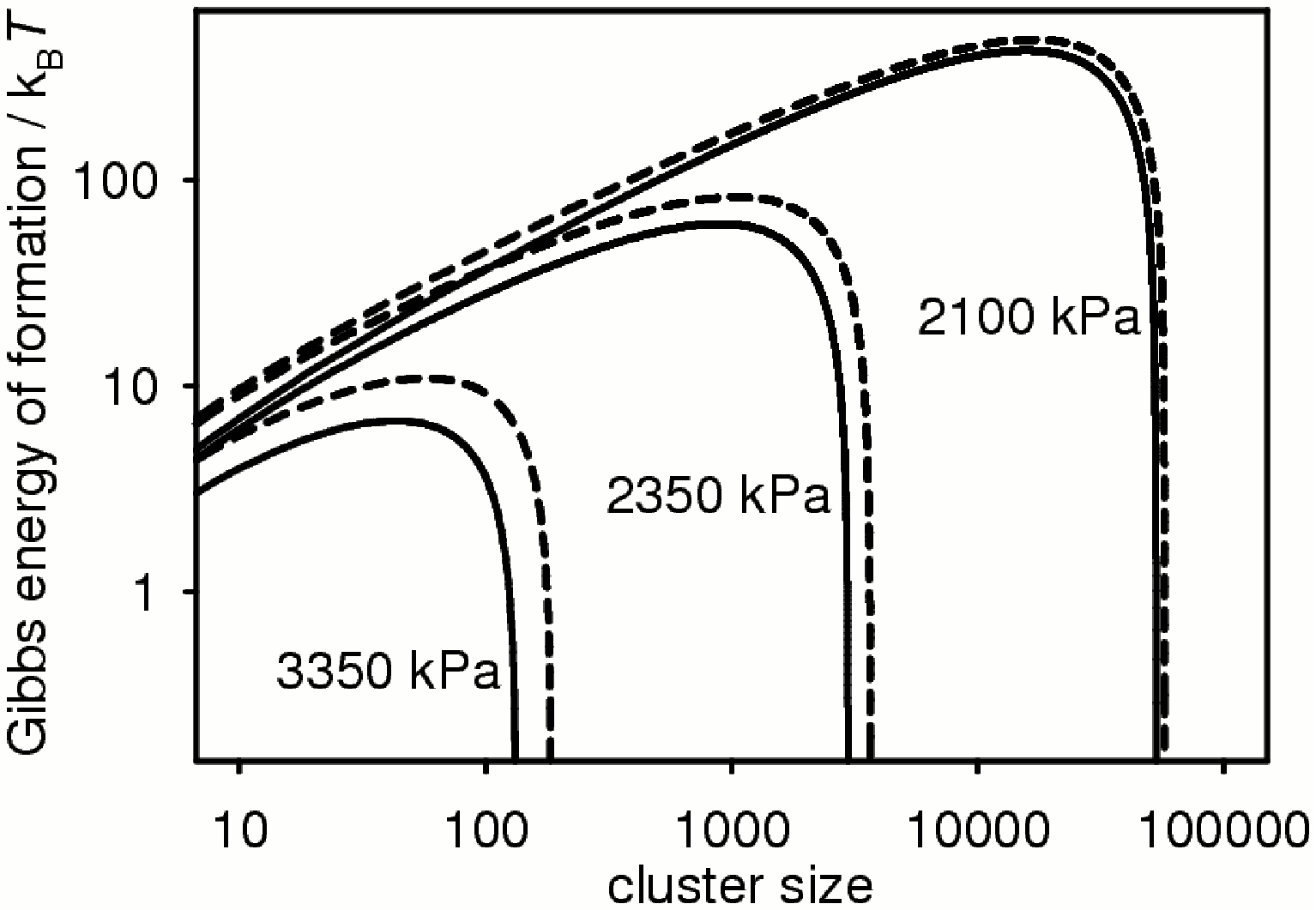}
      \quad\quad
   \includegraphics[width=5.67cm]{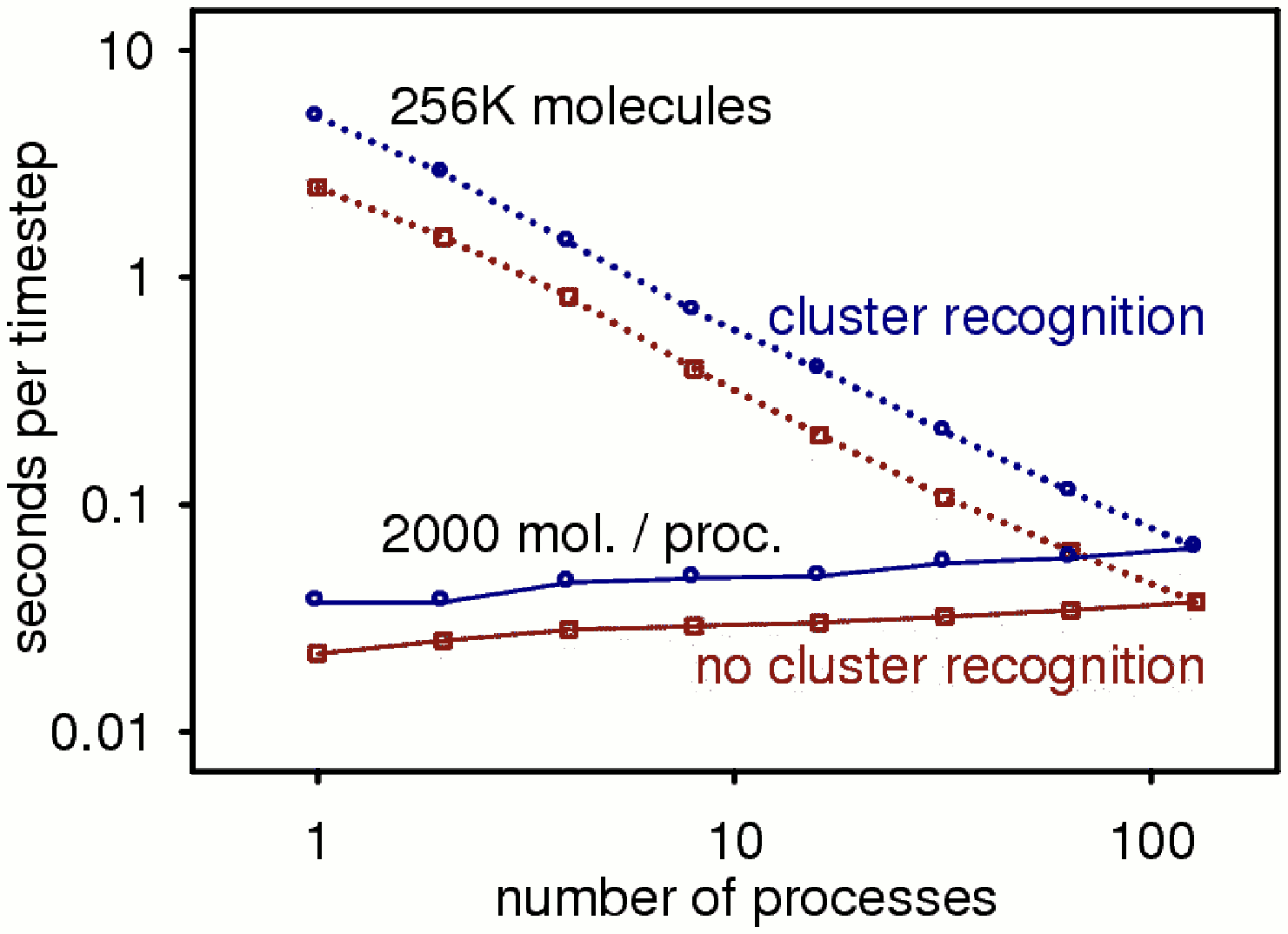}
      \quad\quad
   \includegraphics[width=5.67cm]{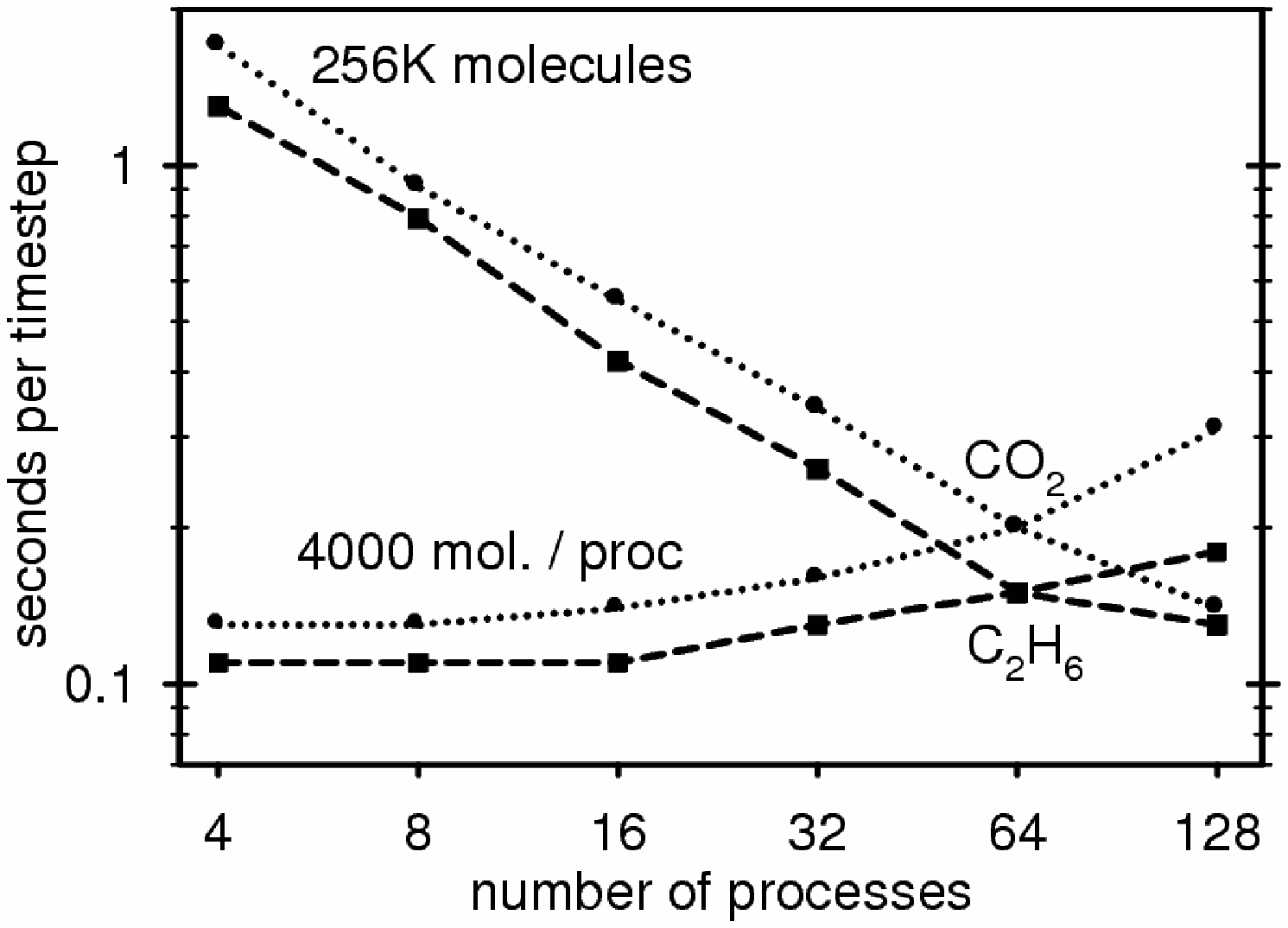}
}
\caption{
   \textbf{left} \,--\,
      Gibbs energy of cluster formation in CO$_2$ at 253 K for pressure values of 2100,
      2350, and 3350 kPa according to CNT (\,---\,) and LFK (\,-\,-\,-\,);
   \quad \textbf{center} \,--\,
      Strong (256000 molecules) and weak scaling (2000 molecules per process) of \lsone{}
      simulating CH$_4$ at 102 K and 0.730 mol/l on the Intel Xeon cluster \textit{mozart}
      at the chairs \textit{Simulation of Large Systems} and \textit{Numerics for Supercomputers},
      Universit\"at Stuttgart;
   \quad \textbf{right} \,--\,
      Strong (256000 molecules) and weak scaling (4000 molecules per process)
      of \lsone{} for 2CLJQ fluid models on the Intel Xeon cluster \textit{cacau}
      at the \textit{High Performance Computing Center Stuttgart}: C$_2$H$_6$ at 183 K and
      0.365 mol/l (\,-\,-\,-\,) as well as CO$_2$ at 253 K and 1.670 mol/l (\,$\cdots$\,)
}
\label{alephI}
\label{alephII}
\label{alephIII}
\end{figure}

\section{\bf\large Nucleation theories}
\label{sec:theories}

\begin{multicols}{2}
\subsection{Classical nucleation theory}

The foundations of CNT were laid by Gibbs \cite{Gibbs28} and further
developed by Volmer and Weber \cite{VW26}. Important subsequent contributions were made
by Farkas \cite{Farkas27}, Becker and D\"oring \cite{BD35},
Zel'dovich \cite{Zeldovich42}, and Feder et al. \cite{FRLP66}.
For the further development of the theory
compare Kashchiev \cite{Kashchiev00} and Vehkam\"aki \cite{Vehkamaeki06}.

The starting point of this theory is the capillarity approximation: 
the dispersed liquid phase, composed of the clusters emerging during nucleation,
is assumed to have the same thermodynamic properties as the saturated bulk liquid.
It is also assumed that all liquid clusters are spherical.
CNT describes how, under such preconditions, nucleation rate and critical size
depend on temperature, supersaturation, and a few properties of the fluid
which are independent of $\supersat$, such as the planar interface
surface tension $\planartension$ and the densities $\satnumdensity$ and $\liqnumdensity$,
referring to the saturated vapor and liquid, repsectively.

The surface energy $\surfaceenergy(\csizea)$ of a cluster with $\csizea$ molecules,
the surface area $\area(\csizea)$, and the specific surface energy $\specenergy$ amounts
to $\specenergy\area(\csizea)$. The capillarity approximation assigns $\specenergy = \planartension$,
and CNT further assumes spherical clusters, hence
$\,\area(\csizea) \,=\, \sqrt[3]{\pi} \, {(6\csizea\slash\liqnumdensity})^{2\slash{}3}$.
The Gibbs energy of cluster formation in a supersaturated vapor consists of a negative bulk
contribution and a positive surface contribution \cite{Gibbs28, VW26}. It amounts to
\begin{equation}
\label{eqnWOF}
\Delta\gibbsenergy_\csizea \,\,=\,\, \surfaceenergy(\csizea) - \surfaceenergy(1) + (1 - \csizea)\Delta\chempot,
\end{equation}
and reaches a maximum at
the size $\critsize$ of the critical nucleus.
It can be seen from Figure \ref{alephI} (left) that the critical size is strongly dependent
on the supersaturated vapor pressure; it diverges as the supersaturated vapor pressure $\pressure$
approaches the saturated vapor pressure $\psat$ of the bulk fluid.
The chemical potential difference,
\begin{equation}
\Delta\chempot \,=\, \int_{\psat}^{\pressure} d\pressure\slash\vapnumdensity,
\label{eqn:deltamu}
\end{equation}
is an integral between $\psat$ and $\pressure$ at constant temperature.
In metastable equilibrium,
the $\csizea$-clus\-ter number density $\numdensity_\csizea = \absnum_\csizea \slash \volume$,
where $\absnum_\csizea$ is the number of clusters with exactly $\csizea$ molecules, amounts to
\begin{equation}
   \numdensity_\csizea \,\,=\,\,
      \numdensity_1 \exp\left(\frac{-\Delta\gibbsenergy_\csizea}{\kboltz\temperature}\right),
   \label{eqnrho}
\end{equation}
where $\numdensity_1$ can be estimated from
\begin{equation}
   \numdensity \,\,\simeq\,\, \sum_{\csizea = 1}^\critsize \csizea\numdensity_\csizea.
   \label{eqnrhoone}
\end{equation}
The impact rate $\impactrate$ of vapor molecules on a cluster per surface area can be approximated by
\begin{equation}
   \impactrate \,\,=\,\, \frac{\pressure}{\sqrt{2\pi\molecularmass\kboltz\temperature}},
\end{equation}
where $\molecularmass$ is the molecular mass \cite{HP63}.
Assuming that every collision of a monomer with a critical nucleus leads to the formation of a cluster
with $1 + \critsize$ molecules, the nucleation rate is given by
\begin{equation}
   \nuclrate \,\,=\,\, \critnucldensity\impactrate\critnuclarea\zeldovich\noniso.
   \label{eqnJ}
\end{equation}
Here and elsewhere, all quantities marked with an asterisk refer to critical nuclei.
The factor $\impactrate\critnuclarea$ expresses the impact frequency of monomers on a critical nucleus,
or equivalently, the rate at which critical nuclei grow to a size of $1+\critsize$ molecules.

The two remaining factors, $\zeldovich$ and $\noniso$, represent corrections with respect to
the nucleus density, the kinetics of the nucleation process, and the temperature inside a nucleus.
The metastable equilibrium breaks down near the critical size, and the actual number density of
critical nuclei is considerably lower than their metastable equilibrium density $\critnucldensity$.
The Zel'dovich factor,
\begin{equation}
   \zeldovich \,\,=\,\, \sqrt{\frac{-1}{2\pi\kboltz\temperature}
      \left(\frac{\partial^2\gibbsenergy_\csizea}{\partial\csizea^2}\right)_\critsize}
         \,\,=\,\, \frac{1}{3\critsize}\sqrt{\frac{\surfaceenergy^*}{\pi\kboltz\temperature}},
\end{equation}
takes into account both this difference and the probability that
that a nucleus above the critical size does not continue to grow \cite{Zeldovich42}.

Nuclei that reach the critical size usually have grown very fast and retain part
of the latent heat.
Let $\cv$ be the isochoric heat capacity of the vapor and $\enthalpy$ the enthalpy of
evaporation. From considerations of Feder et al.\ \cite{FRLP66} it follows that
on average, critical nuclei are overheated by
\begin{equation}
   \DeltaTcrit = \frac{2\kboltz\temperature^2\zeldovich}{\enthalpy}.
\label{eqnDeltaTcrit}
\end{equation}
This increase in temperature reduces the nucleation rate, an effect
which is quantified by the non-isothermal factor $\noniso$.
The energy added to a critical nucleus when it acquires an additional molecule is
\begin{equation}
   \releaseenergy = \enthalpy - \frac{\kboltz\temperature}{2}
      - \left(\frac{\partial\surfaceenergy(\csizea)}{\partial\csizea}\right)_\critsize
         = \enthalpy - \frac{\kboltz\temperature}{2} - \frac{2\surfaceenergy^*}{3\critsize},
\end{equation}
in excess of what is needed to extend its area at the same temperature.
The standard deviation of the energy of vapor molecules that collide with a cluster is
\begin{equation}
   \energydeviation \,\,=\,\, \temperature \sqrt{\kboltz(\cv + \kboltz\slash2)}.
\end{equation}
Finally, the non-isothermal factor is given by
\begin{equation}
   \noniso \,\,=\,\, \frac{\energydeviation^2}{\energydeviation^2 + \releaseenergy^2}.
\end{equation}

\subsection{Model proposed by Laaksonen, Ford, and Kulmala}

The LFK model \cite{LFK94} is a version of the Dillmann-Meier approach \cite{DM90}
which postulates the surface energy of a cluster with $\csizea \in \naturals$ molecules to be
\begin{equation}
   \surfaceenergy(\csizea) \,\,=\,\, \lfkK(\csizea)\planartension\area(\csizea)
      + \lfktau\kboltz\temperature\ln\csizea.
   \label{eqnDM}
\end{equation}
The adjustable parameters of this model are $\lfktau$ and $\lfkK(\csizea)$ for $\csizea \in \naturals$,
as well as $\numdensity_1$ which is expressed indirectly by means of a normalization parameter $\lfkqzero$.
By comparing the Fisher \cite{Fisher67} equation of state,
\begin{equation}
   \pressure \,\,=\,\, \kboltz\temperature\sum_{\csizea\in\naturals} \numdensity_\csizea,
\end{equation}
to a virial-type expansion of second order
values for $\lfkK(1)$ and $\lfkK(2)$ are defined. Laaksonen et al.\ \cite{LFK94} represent this
in terms of the monomer density as
\begin{eqnarray}
   \numdensity_1 \,\,=\,\, \frac{\pressure}{\kboltz\temperature}
                           \left(1 + \frac{\virial\pressure}{\kboltz\temperature}\right)
                 \,\,=\,\, \frac{\pressure^2}{\numdensity(\kboltz\temperature)^2},
\end{eqnarray}
where the second virial coefficient is, in this case, defined as
\begin{equation}
   \virial \,\,=\,\, \numdensity^{-1} - \pressure^{-1}\kboltz\temperature.
   \label{eqn:virialLFK}
\end{equation}
They obtain the expressions
\begin{eqnarray*}
   \lfkK(1) &=& -\frac{1}{\lfkTheta} 
      \left(\ln\left(\frac{\psat}{\lfkqzero\kboltz\temperature}\right)
         + \frac{\virial\psat}{\kboltz\temperature}\right), \\
   \lfkK(2) &=& -\frac{1}{\lfkTheta2^{2\slash3}}
      \left(\frac{\virial\psat}{\kboltz\temperature} - \lfkK(1)\lfkTheta +
         \ln\left(\frac{-2^\tau\virial\psat}{\kboltz\temperature}\right)\right),
\end{eqnarray*}
with $\lfkTheta = \planartension\area(1) \slash \kboltz\temperature$, by applying
an approximation to Equation (\ref{eqnrhoone}). This is extended to higher $\lfkK(\csizea)$ by
\begin{equation}
   \lfkK(\csizea) \,\,=\,\, 1 + \lfkalpha_1\csizea^{-1\slash3}
                              + \lfkalpha_2\csizea^{-2\slash3},
   \label{eqnalpha}
\end{equation}
which are the first three contributions of an expansion in terms of the inverse cluster radius
$\clusterradius^{-1} \sim \csizea^{-1\slash3}$. The coefficients $\lfkalpha_1$ and $\lfkalpha_2$
are determined by equating the expressions for $\lfkK(1)$ and $\lfkK(2)$ with
Equation (\ref{eqnalpha}).

Note that since Equation (\ref{eqnDM}) multiplies $\lfkK(\csizea)$
with $\area(\csizea) \sim \csizea^{2\slash3}$, the value of $\lfkalpha_2$ only influences a constant
summand which cancels out in the expression for $\Delta\gibbsenergy_\csizea$.
Laaksonen et al.\ \cite{LFK94} proposed $\lfktau = 0$, and Ford et al.\ \cite{FLK93}
showed that with this particular assignment, the parameter $\lfkqzero$
cancels out as well. To obtain a convenient expression,
we set $\lfkqzero = \psat\slash\kboltz\temperature$, which leads to
\begin{eqnarray}
   \lfkK(1) &\,\,=\,\,& \frac{-\virial\psat}{\planartension\area(1)}, \\
   \lfkK(2) &\,\,=\,\,& \frac{-2\virial\psat
      - \kboltz\temperature\ln(-\virial\psat \slash \kboltz\temperature)}
         {\planartension\area(2)}.
\end{eqnarray}
The Zel'dovich factor takes the form
\begin{equation}
   \zeldovich \,\,=\,\, \frac{1}{3\critsize}\sqrt{
      \frac{\planartension\area(1)}{\pi\kboltz\temperature}
         \sqrt[3]{\critsize}\left(\lfkalpha_1 + \sqrt[3]{\critsize}\right)},
\end{equation}
and the energy released on addition of a monomer to a cluster amounts to
\begin{equation}
   \releaseenergy \,\,=\,\, \enthalpy - \frac{\kboltz\temperature}{2}
      - \frac{\planartension\area(1)}{3\sqrt[3]{\csizea}}\left(2 + \frac{\lfkalpha_1}{\sqrt[3]{\csizea}}\right).
\end{equation}
Compared to CNT, the LFK model hence introduces a size dependence of the specific surface
energy which is governed by the single parameter $\lfkalpha_1$. Figure \ref{alephI} (left)
illustrates that this size dependence becomes particularly significant at high supersaturations,
where the critical nucleus is small.

\subsection{Mean first passage times}

Let us next consider the kinetics of a nucleation process.
For a supersaturated fluid in a volume $\volume$ that exhibits the nucleation rate $\nuclrate$, it might
be expected that the first stable nuclei appear on average after a temporal delay,
expressed by
\begin{equation}
\mfpt(\csizea) \,\,\approx\,\, \frac{1}{\nuclrate\volume},
\label{eqnmfptidea}
\end{equation}
for some $\csizea > \critsize$,
after the onset of nucleation \cite{Wedekind06}. The average delay $\mfpt(\csizea)$ until the first cluster with
$\csizea$ molecules appears is called the {\em mean first passage time} of $\csizea$.
Wedekind et al.\ \cite{WSR07, WRS06, WWRS07} generalized this
approach to a theory of condensation processes, here referred to as MFPT.
Bartell and Wu \cite{BW06} obtained an identical result for freezing,
and Zhang et al.\ \cite{ZAXSL07} applied it to melting processes.
According to this approach, the mean first passage time is approximated using a Gauss error function,
\begin{equation}
\mfpt(\csizea) \,\,=\,\, \frac{\mfpt_\infty}{2}\,\left[\,1 + \erf\mathbf{(}\,\reala(\csizea - \critsize)\,\mathbf{)}\,\right].
\label{eqn:mfpt}
\end{equation}
In particular, this approach leads to
\begin{equation}
\lim_{\csizea\to\infty} \mfpt(\csizea) \,\,=\,\, \mfpt_\infty \,\,=\,\, 2\mfpt^*,
\label{eqn:mfpt2}
\end{equation}
and thus
\begin{equation}
\nuclrate \,\,\approx\,\, \frac{1}{2\mfpt^*\volume},
\label{eqn:mfpt3}
\end{equation}
from Equation (\ref{eqnmfptidea}) with $\csizea \to \infty$ \cite{Wedekind06}.
These approximations are intended to hold only \qq{in the vicinity of the critical size} and 
\qq{[u]nder reasonably high barriers} \cite{WSR07}.
%
\end{multicols}

\begin{table}[h]\centering
\begin{tabular}{l|l|rrrrr}
\hline\hline
 & model & $\molecularmass$ [u] & $\LJsigma$ [\angstrom] & $\LJepsilon$ [$\kboltz \,\times$ K]
 & $\quadrupole$ [B] & $\elongation$ [\angstrom] \\ \hline 
CH$_4$ & LJ & 16.04 & 3.7281 & 148.55 & & \\
C$_2$H$_6$ & 2CLJQ & 2 $\times$ 15.03 & 3.4896 & 136.99 & 0.8277 & 2.3762 \\
CO$_2$ & 2CLJQ & 2 $\times$ 22.00 & 2.9847 & 133.22 & 3.7938 & 2.4176 \\
\hline\hline
\end{tabular}
\caption{Parameters of the molecular models for methane, ethane, and carbon dioxide}
\label{tabmodels}
\end{table}

\begin{figure}[h]
\centerline{
   \includegraphics[width=5.67cm]{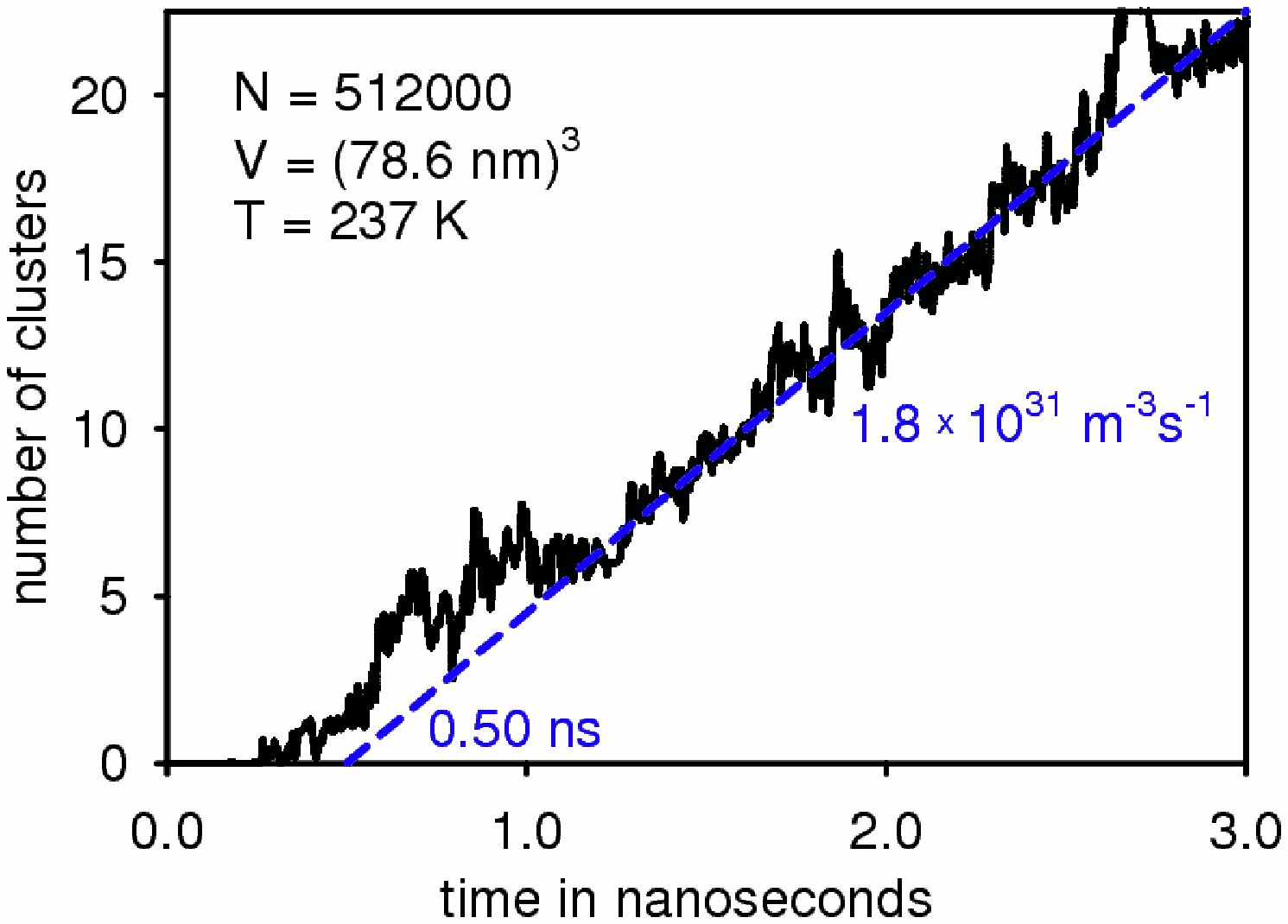}
      \quad\quad
   \includegraphics[width=5.67cm]{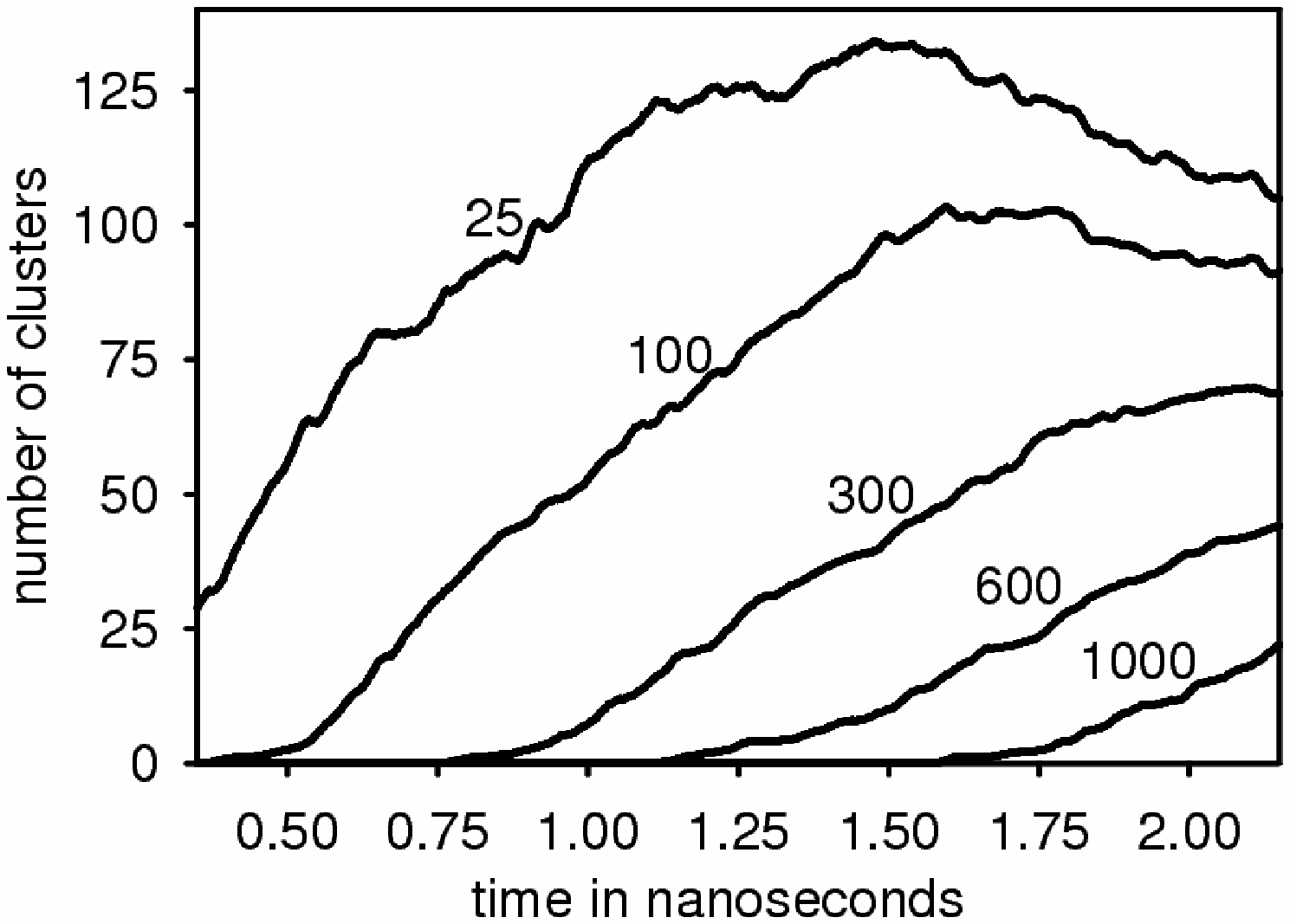}
      \quad\quad
   \includegraphics[width=5.67cm]{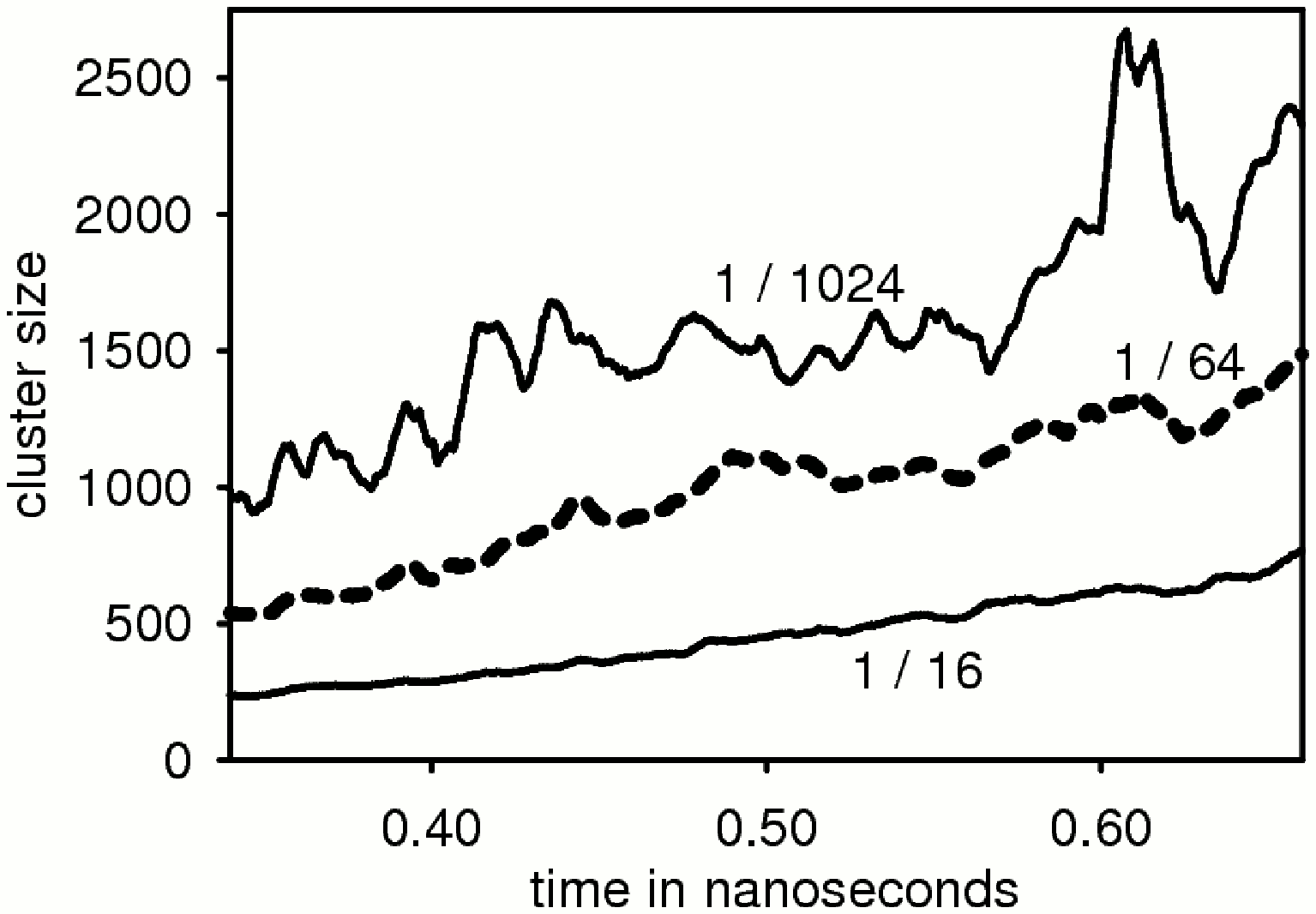}
}
\caption{
   \textbf{left} \,--\,
      Number of nuclei containing at least 75 molecules in supersaturated CO$_2$ vapor over simulation time;
   \quad\textbf{center} \,--\,
      Number of nuclei containing at least 25, 100, \dots, 1000 molecules over simulation time in
      (63.7 nm)$^3$ filled with methane at 130 K and 1.606 mol/l;
   \quad\textbf{right} \,--\,
      Cluster formation delay $\frontier_\reala(\timea)$ for CO$_2$ at 253 K and 3.150 mol/l
      with $\absnum = 884700$ and $\reala \,\in\, \{2^{-4}, 2^{-6}, 2^{-10}\}$.
}
\label{alephXIII}
\label{alephXIV}
\label{alephXI}
\end{figure}

\section{\bf\large Simulation method}
\label{sec:sim}
\begin{multicols}{2}
Methane, ethane and carbon dioxide were selected in the present work because of their qualitatively different
molecular properties. Methane is almost spherical and weakly octupolar, thus it can be described by a single
Lennard-Jones site with the pair potential
\begin{equation}
   \potentialLJ(\distance_{\inta\intb}) \,\,=\,\, 4\LJepsilon \, \left( \,
      \left(\frac{\LJsigma}{\distance_{\inta\intb}}\right)^{12} -
      \left(\frac{\LJsigma}{\distance_{\inta\intb}}\right)^{6} \, \right).
\end{equation}
Ethane molecules are dumbbell-shaped and thus significantly anisotropic in geometry but only slightly quadrupolar.
Carbon dioxide molecules are both strongly
anisotropic and quadrupolar. The intermolecular interactions of these two fluids were described by the
two-center LJ model with an embedded point quadrupole (2CLJQ). Additional parameters of the 2CLJQ model
are the molecular elongation $\elongation$ and the quadrupole moment $\quadrupole$.
The parameters of the molecular models, cf.\ Table \ref{tabmodels}, were adjusted to
experimental vapor-liquid equilibria in prior work \cite{VSH01}.

%
%
Series of MD simulations of nucleating vapors were conducted using a version of
the \lsone{} program \cite{BV05}. The simulations were carried out in the canonical ensemble, 
with a time step between 3 and 7 fs, depending on the system temperature.
The cutoff radius $\distancecut$ was larger than $4.5 \LJsigma$ in all simulations.
The temperature of the whole system was kept
constant by isokinetic scaling.

To follow the kinetics of the phase transition in detail,
a criterion which detects clusters of molecules, i.e.\ the dispersed liquid phase,
must be applied to the whole ensemble.
In past studies, a considerable number of different cluster criteria were
discussed and compared \cite{BM01, Barrett02, Sator03, GRVE07, WR07}.
Those presented by Hill \cite{Hill55} and Stillinger \cite{Stillinger63}
are among the most common ones. They are applied
to all pair interactions and the clusters are defined as the connected components
of the graph with the molecules as its nodes and edges between the pairs with interactions
that fulfill a pair critierion. The Hill energetic criterion is defined by
\begin{equation}
   \internal(\distance_{\inta\intb})
      + \frac{\molecularmass\velocity_\textnormal{\tiny{}rel}^2}{2} \,\,<\,\, 0,
\end{equation}
and the Stillinger geometric criterion by
\begin{equation}
  \distance_{\inta\intb} \,\,<\,\, \stillingerdistance,
\end{equation}
where $\stillingerdistance = 1.5\LJsigma$ for the Lennard-Jones fluid.
The above definitions distinguish the bulk phases.
%
%
A hybrid cluster criterion which combines these definitions
was consistently observed to select only few clusters with extremely short
lifetimes, whereas it reliably detected stable clusters of all sizes. It is defined as follows:
\begin{itemize}
   \item All molecules $\inta$ for which the energetic single-molecule criterion
         \begin{equation}
            \molecularmass\velocity_\inta^2 +
            \Sigma_{\inta\neq\intb} \internal_\textnormal{\tiny{}pot}(\distance_{\inta\intb}) \,\,<\,\, 0,
         \end{equation}
         holds are defined to be \textit{liquid}.
   \item Two liquid molecules are regarded as \textit{connected} whenever they fulfill the Stillinger criterion.
         For the 2CLJQ model the maximal connection radius is given by 
         \begin{equation}
            \stillingerdistance \,\,=\,\, \frac{3\LJsigma}{2} + \frac{\elongation}{4}.
         \end{equation}
   \item Clusters are determined by covering a graph consisting of these connections with
         maximal \textit{biconnected components} and eliminating their overlap.
         Monomeric clusters are regarded as vapor molecules.
\end{itemize} 
Biconnected components are, by definition, subsets of a graph that cannot be separated into two unconnected parts
by removing only one vertex. This reflects the idea that a cluster should not consist of several sub-nuclei
connected by a single molecule, because structures that do depend on such a connection tend to be extremely
unstable.

\begin{figure}[h]
\centerline{
   \includegraphics[width=5.67cm]{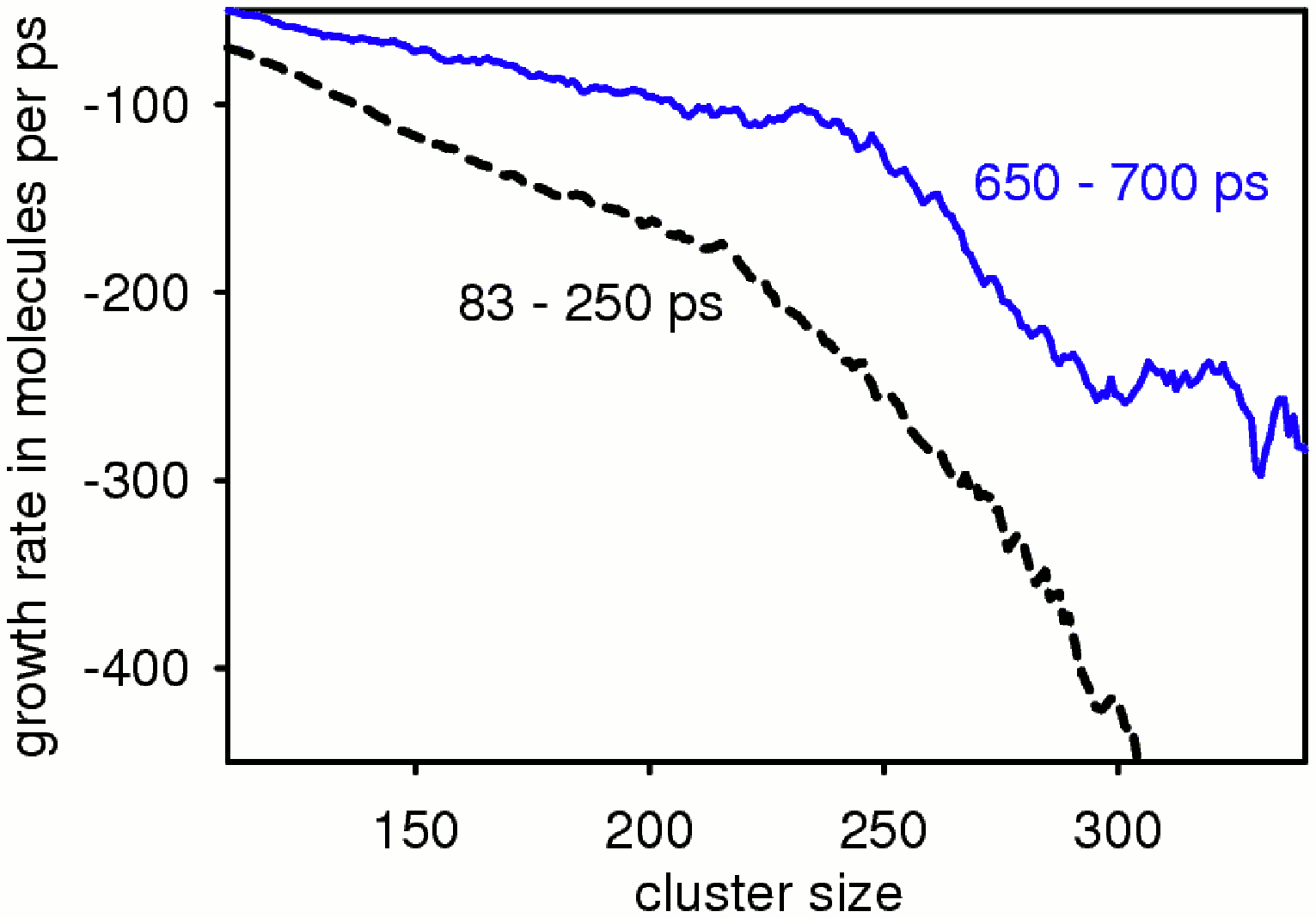}
      \quad\quad
   \includegraphics[width=5.67cm]{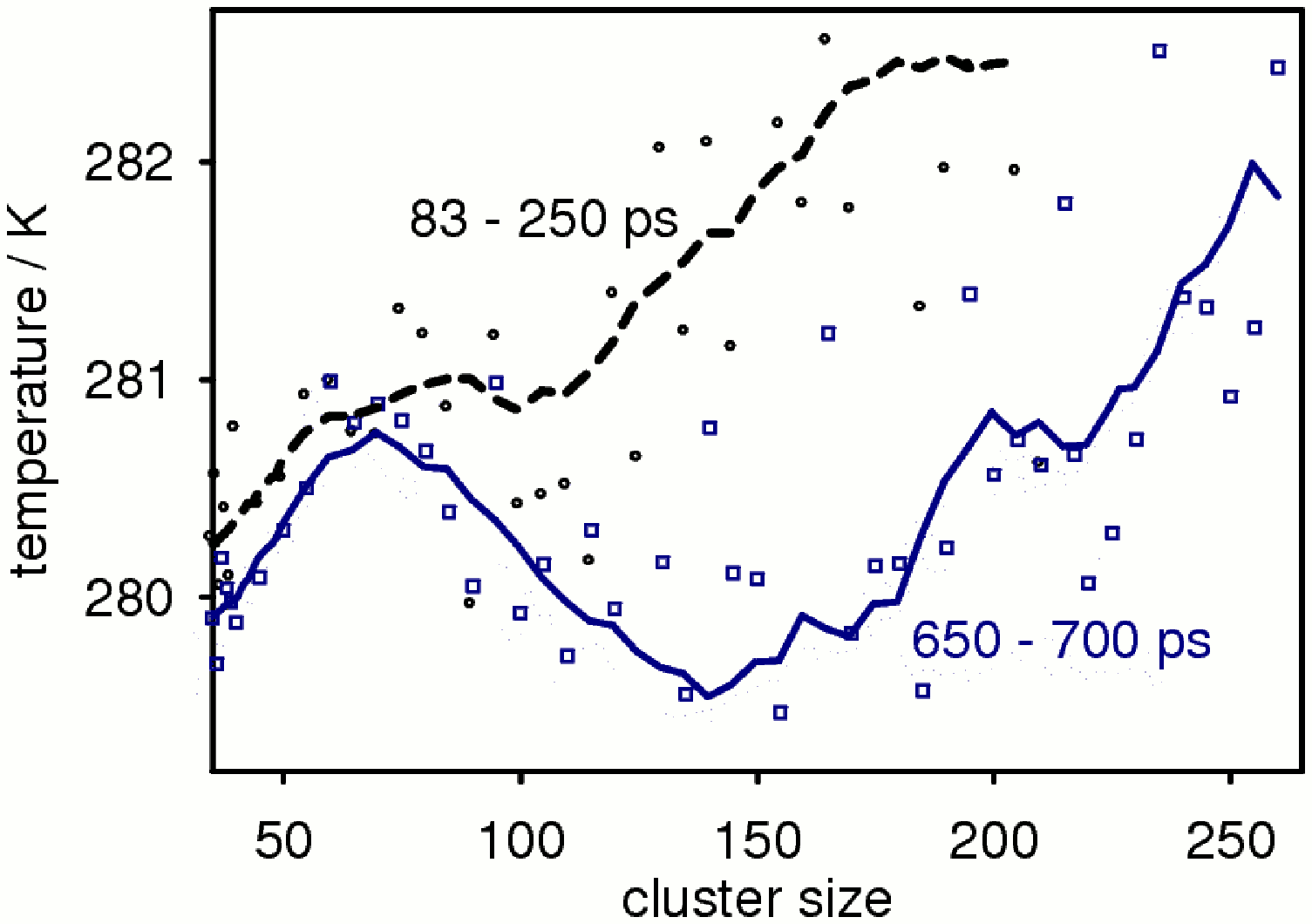}
      \quad\quad
   \includegraphics[width=5.67cm]{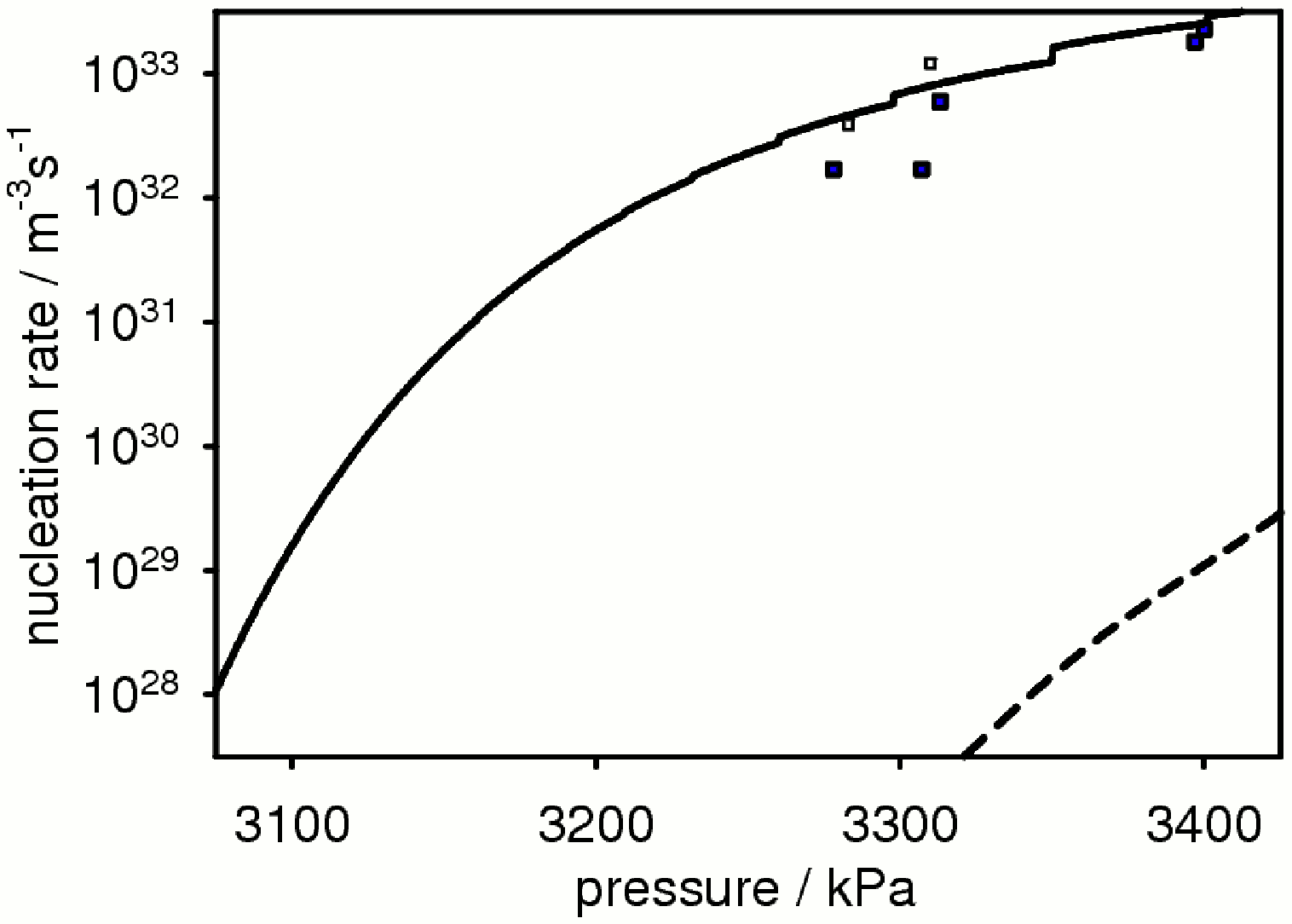}
}
\caption{
   \textbf{left} \,--\,
      Cluster net growth rate as a function of cluster size and time in C$_2$H$_6$ at 280 K and
      2.800 mol/l; 
   \quad\textbf{center} \,--\,
      Cluster temperature as a function of cluster size and time in C$_2$H$_6$ at 280 K and
      2.800 mol/l; data (circles) and running average (\,-\,-\,-\,) from 83 to 250 ps as well as data (squares)
      and running average (\,---\,) from 650 to 700 ps after the initial state;
   \quad\textbf{right} \,--\,
      Nucleation rate of C$_2$H$_6$ at 280 K; small squares: \nuclrate(50),
      large squares: \nuclrate(75) and \nuclrate(100)
}
\label{alephVI}
\label{alephVII}
\label{alephIX}
\end{figure}

The Hill energetic criterion favors molecules with a low kinetic energy, and hence leads
to artefacts in the cluster temperature, i.e.\ clusters are observed to be colder than they actually are.
This effect carries over to the hybrid criterion.
For this reason, temperature data as displayed in Figure \ref{alephVII} (center) were gathered by applying
only the geometric and the biconnectivity parts of the hybrid criterion.

The MD program \lsone{} relies on spatial domain decomposition for parallel simulations \cite{BV05}.
The operation of partitioning a very large graph into biconnected components
was handled by including the \boost{} library \cite{SLL02} which
implements Tarjan's linear time algorithm.
Figure \ref{alephII} (center and right) shows that \lsone{}, both with and without cluster recognition,
scales well on typical clusters of workstations.
This permits simulations of volumes $\volume \approx 10^{-21}\, \textnormal{m}^3$ for a time span
of a few nanoseconds with an acceptable computational effort. Thus with the direct
approach, which requires at least some stable nuclei to appear, only values
$\nuclrate > 10^{30} / (\textnormal{m}^3\textnormal{s})$ are accessible unless correspondingly
larger computational resources are employed. 
\end{multicols}

\begin{table}[h] \centering
\begin{tabular}{l|cccccc}
\hline\hline
 & \temperature \, [\,K\,] & \numdensity \, [\,mol\slash{}l\,]
 & MFPT \nuclrate  & direct MD $\nuclrate(10)$ \\ \hline 
Ar & 50.0 & 0.139 & $1\times{}10^{31}$ \slash{}(m$^3$s) & $5\times{}10^{30}$ \slash{}(m$^3$s) \\
CH$_4$ & 63.6 & 0.105 & $1\times{}10^{31}$ \slash{}(m$^3$s) & $6\times{}10^{30}$ \slash{}(m$^3$s) \\
\hline\hline
\end{tabular}
\label{tabmfpt}
\caption{Comparison of the nuclation rate from an MFPT indirect analysis according to Wedekind et al.\ \cite{WRS06}
with the value $\nuclrate(10)$ from another simulation of that system analyzed with the method of Yasuoka and Matsumoto;
a single set of values for the LJ fluid is interpreted as both argon and methane}
\end{table}

\begin{table}[h] \centering
\begin{tabular}{|rrr|rr|rrrr|}
\hline\hline
   $\superrho$ & $\superp$
 & $\supermu$
 & $\csizea$ & $\nuclrate(\csizea)$ [m$^{-3}$s$^{-1}$]
 & $\critsize(\textnormal{CNT})$
 & $\nuclrate(\textnormal{CNT})$ [m$^{-3}$s$^{-1}$]
 & $\critsize(\textnormal{LFK})$
 & $\nuclrate(\textnormal{LFK})$ [m$^{-3}$s$^{-1}$] \\
\hline

   $1.694$ & $1.269$ & $1.164$
 & $25$ & $1.3 \times \,10^{33}\,$ & $49$
 & $1.1 \times \,10^{34}\,$ & $132$ & $4.1 \times \,10^{30}\,$ \\
   $1.694$ & $1.263$ & $1.161$
 & $75$ & $\mathbf{1.8 \times 10^{32}}$ & $51$
 & $\mathbf{8.0 \times 10^{33}}$ & $139$ & $2.5 \times \,10^{30}\,$ \\
   $1.694$ & $1.250$ & $1.154$
 & $225$ & $\mathbf{1.5 \times 10^{32}}$ & $57$
 & $\mathbf{6.6 \times 10^{33}}$ & $153$ & $\mathbf{8.7 \times 10^{29}}$ \\

   $1.769$ & $1.273$ & $1.166$
 & $25$ & $2.1\times \,10^{33}\,$ & $47$
 & $1.2 \times \,10^{34}\,$ & $128$ & $6.2 \times \,10^{30}\,$ \\
   $1.769$ & $1.259$ & $1.159$
 & $75$ & $\mathbf{4.0 \times 10^{32}}$ & $53$
 & $\mathbf{7.6 \times 10^{33}}$ & $143$ & $1.8 \,\times 10^{30}\,$ \\
   $1.769$ & $1.247$ & $1.153$
 & $225$ & $\mathbf{2.7 \times 10^{32}}$ & $59$
 & $\mathbf{5.2 \times 10^{33}}$ & $157$ & $\mathbf{6.7 \times 10^{29}}$ \\

\hline\hline
\end{tabular}
\caption{Simulation results and theoretical values of nucleation rates for supersaturated
methane at 170 K with \psat = 2328 kPa, \ndsat = 2.429 mol\slash{}l, and \planartension = 2.07 g/s$^2$;
bold values: threshold $\iota > \critsize$ according to theory}
\label{tabCH4T170K}
\end{table}

\begin{table}[h] \centering
\begin{tabular}{|rrr|rr|rrrr|}
\hline\hline
   $\superrho$ & $\superp$
 & $\supermu$
 & $\csizea$ & $\nuclrate(\csizea)$ [m$^{-3}$s$^{-1}$]
 & $\critsize(\textnormal{CNT})$
 & $\nuclrate(\textnormal{CNT})$ [m$^{-3}$s$^{-1}$]
 & $\critsize(\textnormal{LFK})$
 & $\nuclrate(\textnormal{LFK})$ [m$^{-3}$s$^{-1}$] \\
\hline

   $2.032$ & $1.61$ & $1.42$
 & $50$ & $9.1 \times \,10^{31}\,$ & $58$
 & $1.3 \times \,10^{32}\,$ & $80$ & $8.4 \times \,10^{29}\,$ \\

   $2.096$ & $1.63$ & $1.43$
 & $25$ & $6.2 \times \,10^{31}\,$ & $54$
 & $2.0 \times \,10^{32}\,$ & $75$ & $1.7 \times \,10^{30}\,$ \\

   $2.352$ & $1.70$ & $1.47$
 & $25$ & $6.7 \times \,10^{32}\,$ & $44$
 & $6.9 \times \,10^{32}\,$ & $59$ & $2.5 \times \,10^{31}\,$ \\
   $2.352$ & $1.69$ & $1.46$
 & $100$ & $\mathbf{1.2 \times 10^{32}}$ & $46$
 & $\mathbf{5.9 \times 10^{32}}$ & $61$ & $\mathbf{1.7 \times 10^{31}}$ \\

   $2.481$ & $1.72$ & $1.48$
 & $25$ & $1.4 \times \,10^{33}\,$ & $42$
 & $9.1 \times \,10^{32}\,$ & $55$ & $6.1 \times \,10^{31}\,$ \\
   $2.481$ & $1.70$ & $1.47$
 & $100$ & $\mathbf{2.9 \times 10^{32}}$ & $44$
 & $\mathbf{6.9 \times 10^{32}}$ & $59$ & $\mathbf{2.5 \times 10^{31}}$ \\

   $2.609$ & $1.73$ & $1.48$
 & $25$ & $2.2 \times \,10^{33}\,$ & $41$
 & $1.0 \times \,10^{33}\,$ & $54$ & $6.7 \times \,10^{31}\,$ \\
   $2.609$ & $1.70$ & $1.47$
 & $100$ & $\mathbf{7.6 \times 10^{32}}$ & $44$
 & $\mathbf{6.9 \times 10^{32}}$ & $59$ & $\mathbf{2.5 \times 10^{31}}$ \\

   $2.695$ & $1.71$ & $1.47$
 & $200$ & $\mathbf{8.4 \times 10^{32}}$ & $43$
 & $\mathbf{8.0 \times 10^{32}}$ & $57$ & $\mathbf{3.8 \times 10^{31}}$ \\
   $2.695$ & $1.70$ & $1.47$
 & $500$ & $\mathbf{4.9 \times 10^{32}}$ & $44$
 & $\mathbf{6.9 \times 10^{32}}$ & $59$ & $\mathbf{2.5 \times 10^{31}}$ \\

\hline\hline
\end{tabular}
\caption{Simulation results and theoretical values of nucleation rates for supersaturated
carbon dioxide at 253 K with \psat = 1961 kPa, \ndsat = 1.169 mol\slash{}l, and \planartension = 86.2 g/s$^2$;
bold values: threshold $\iota > \critsize$ according to theory}
\label{tabCO2T253K}
\end{table}

\section{\bf\large Simulation results}
\label{sec:simresults}

\begin{multicols}{2}
\subsection{Growth rates of single nuclei}
\label{sec:singlecluster}

Both CNT and the Dillmann-Meier model assume that the properties of clusters with a given size depend
only on the temperature and supersaturation of the vapor. 
Hence, one should expect droplets of the same size generated earlier and later during a simulation run to have, on average,
constant temperatures and rates of growth and evaporation.

It is known from MD simulations by Tanumura et al.\ \cite{TYE00} that this does not
necessarily hold in the initial period:
the very first clusters of a given
size have a higher kinetic energy than those which belong to the actual metastable vapor.
This is due to the fact that molecules lose potential energy when they attach to a droplet, which transforms
to kinetic energy. The first large clusters have experienced a relatively
fast growth process and hence retain more of this latent heat.
The present simulations confirm this observation, cf.\ Figure \ref{alephVII} (center), which shows
that the largest existing clusters have a temperature of 282 K and above, while the whole system temperature
is fixed at 280 K.
The lower curve, collected between 650 and 700 after the simulation onset, exhibits
a local maximum of the cluster temperature at a size of about 70 molecules. The temperature of smaller clusters
changes comparatively little over simulation time, i.e.\ with respect to the higher curve,
whereas for larger clusters it decreases considerably.
The temperature of the clusters with $\csizea < 70$ has reached a steady state, but no thermal
equilibrium with the vapor, which implies that these clusters are unstable. For $\csizea > 70$,
no steady state is established and the cluster temperature approaches 280 K.
Hence, the local maximum of the temperature plot marks the transition
between unstable clusters and stable nano-droplets, i.e.\ the size of the critical nucleus. It
agrees well with the critical size of 78 indicated by CNT, as opposed to the LFK model which
predicts $\critsize = 207$, cf.\ Table \ref{tabseveral}.

As an effect which is closely related to the overheating of the dispersed phase,
clusters generated early in the condensation process evaporate at a higher rate than those which are generated
later, cf.\ Figure \ref{alephVII} (left). These data were collected during the same simulation as those from
Figure \ref{alephVII} (center). Note how the growth rate of clusters is negative for sizes significantly
larger than the critical nucleus, where $\critsize$ can be estimated either from the temperature profile
any of the theories. This phenomenon was also observed by Yasuoka and Matsumoto \cite{YM98}.
The positive contribution to cluster growth, which is due to condensation of the vapor phase,
remains constant over simulation time. This indicates that while the temperature of clusters with a given size changes,
the temporal evolution of the system does not significantly affect their surface area.
From the decrease in cluster evaporation over time, it clearly follows that for MD simulations starting
from a cluster-free configuration, the first clusters differ
significantly from the much larger number of clusters of the same size that appear at a later
stage of the process.

Changing rates of evaporation also imply that the critical size can actually vary during a simulation run
with a very high supersaturation. For systems at very high supersaturations, it is impossible to observe
a metastable vapor phase, because nucleation begins immediately.
Such phenomena are only realistic if it is technically possible to
increase the supersaturation faster than the vapor phase can produce small clusters.

A magic number effect can be observed for small clusters of both the LJ and the 2CLJQ fluids:
the rate of evaporation is comparatively low for clusters with
$\csizea \,\in\, \{8, 11, 14, \dots, 26\}$ molecules. As opposed to this,
clusters with $\csizea \,\in\, \{4, 9, 12, 15\}$ are detected to have particularly high rates of evaporation.
Within this range, 23 and 26, but also 13 and 19, are known as
preferred cluster sizes of the LJ fluid \cite{IHHK96}. 
In the present study, the magic
numbers may well be a side effect of the biconnectivity requirement of the hybrid cluster
criterion. This conclusion is also suggested by the fact that
the observed magic numbers do not depend on the employed molecular model.

\subsection{Nucleation rates}
\label{sec:nuclrate}

A nucleation process {\em at constant pressure} in an infinitely large system occurs, by definition,
with the nucleation rate
\begin{equation}
   \nuclrate \,\,=\,\, \lim_{\csizea \to \infty} \, \lim_{\timea \to \infty} \,
       \frac{d\numdensity_\csizea(\timea)}{d\timea}.
\end{equation}
From molecular simulation in the canonical ensemble, a smoothed $\tilde{\numdensity}_\csizea(\timea)$,
where the statistical noise is reduced, can be
constructed from $\numdensity_\csizea(\timea)$ by averaging over a number of time steps. Such
smoothed plots are shown in Figure \ref{alephXIII} (left and center).
The nucleation rate may then be approximated by the expression
\begin{equation}
   \nuclrate(\csizea) \,\,=\,\, \max_{\timea > \timea_0}
      \frac{d\tilde{\numdensity}_\csizea(\timea)}{d\timea}.
\end{equation}
This approach was introduced by Yasuoka and Matsumoto \cite{YM98}. The values of
$\nuclrate(\csizea)$ are meaningful for all $\csizea \geq \critsize$. However, it has
to be taken into account that as the condensation proceeds in a closed system, the number of monomers
decreases and the pressure in the vapor is reduced significantly,
which causes larger nuclei to be formed at a lower rate, cf.\ Figure \ref{alephXIV} (center).

The present simulation data suggest that, as expected,
the values of $\nuclrate(\csizea)$
are similar for values of $\csizea$ above a certain value, except for cases where the
supersaturation decreases significantly, cf.\ 
Figures \ref{alephIX} (right) and \ref{alephX} (left and center)
as well as Tables \ref{tabCH4T170K}, \ref{tabCO2T253K}, \ref{tabCO2T285K}, and \ref{tabseveral}.
On the other hand, $\nuclrate(\csizea)$ with very small $\csizea$ is often significantly elevated,
which raises doubts whether results related to $\nuclrate(6)$, cf.\ Kraska \cite{Kraska06},
can lead to reliable conclusions.
The nucleation rates are displayed together with pressure values, which
were taken in the center of the interval where the plot of $\tilde{\numdensity}_\csizea(\timea)$
and the linear approximation from which the value of $\nuclrate(\csizea)$ is obtained,
roughly agree -- for instance, after two nanoseconds in the
case of Figure \ref{alephXIII} (left).

Wedekind et al.\ \cite{WSR07} propose an indirect method for the determination of both
the nucleation rate and the critical size from simulation data on mean first passage times.
This approach consists in fitting the values of $\mfpt_\infty$, $\reala$, and $\critsize$
such that Equation (\ref{eqn:mfpt}) agrees optimally with the actual plot of $\mfpt(\csizea)$
from an MD simulation. However, Wedekind et al.\ \cite[Figure 4]{WWRS07} also note that the
size of the critical nucleus determined according to this MFPT based approach can deviate by a factor
larger than two from the \qq{nucleation theorem,}
\begin{equation}
   \frac{\partial\ln\nuclrate}{\partial\ln\superp} \,\,\approx\,\, \critsize + 2,
\label{eqn:nucleationtheorem}
\end{equation}
obtained by Oxtoby and Kashchiev \cite{OK94} in a similar version. That is not necessarily
an argument against the method, since the nucleation theorem is known to
be valid only for moderate supersaturations \cite{Schmelzer01}.
Table \ref{tabmfpt} compares a new MD simulation, evaluated according to the method
of Yasuoka and Matsumoto,
with data obtained by Wedekind et al.\ \cite{WRS06} following the MFPT approach.
The value of $\nuclrate(10)$ is probably larger than $\nuclrate$, since nucleation rates of about
$10^{30}$ -- $10^{31}$ usually imply critical sizes $\critsize \gg 10$.
However, $\nuclrate(10)$ is significantly lower than the MFPT extrapolation.

\subsection{Delay of cluster formation}

Statistics on the formation delay of $\csizea$-clusters are shown in Figure \ref{alephXI} (right).
The plots are of the type
\begin{equation}
   \frontier_\realb(\timea) \,\,\,=\,\,\, \max \, \{\csizeb\in\naturals \,|\,\,
      \Sigma_{(\csizea \,\geq\, \csizeb)}\, \csizea\numdensity_\csizea \,\,\geq\,\, \realb\numdensity\},
\end{equation}
with $0 < \realb \leq 1$, i.e.\ they show the nucleation threshold $\frontier_\realb(\timea)$ passed
by a mole fraction $\realb$ of the condensing fluid at the time $\timea$.

For instance, after
0.5 ns, $\absnum\slash{}16 = 55300$ or more molecules are in clusters with a size
$\csizea \geq \frontier_{1\slash16}(0.5 \textnormal{ns}) = 450$, but the clusters
of $451$ or more molecules contain less than $55300$ molecules. 
At the same time, the threshold corresponding to $\absnum\slash{}1024 = 864$ molecules lies at
$\frontier_{1\slash1024}(0.5 \textnormal{ns}) = 1519$, i.e.\ there are at least 864 molecules in clusters
with $\csizea \geq 1519$, but not in clusters with $\csizea \geq 1520$. Thus the plot
corresponding to $\frontier_{1\slash1024}(t)$ shows the development of the largest cluster.
For that reason, it oscillates more than the other plots.

For $\realb \to 0$, the expected values of $\frontier_\realb(\timea)$ converge by definition towards the inverse
function of $\mfpt(\csizea)$, since by inverting such a plot the first passage times are obtained.
Consider such first passage times from simulations of methane, cf.\ Figure
\ref{alephXII} (right). The data correspond to droplets much larger than the critical nucleus, cf.\ Table
\ref{tabCH4T170K}. From Equation (\ref{eqn:mfpt2}) and (\ref{eqn:mfpt3}) it would be expected that the 
mean first passage time converges according to
\begin{equation}
   \lim_{\csizea\to\infty}\mfpt(\csizea) \,\,=\,\, \frac{1}{\nuclrate\volume},
\end{equation}
which corresponds to 66 ps for 4.116 mol/l and to 38 ps for 4.298 mol/l,
if we accept the values of $\nuclrate(225)$ determined with the method of Yasuoka and Matsumoto.
However, no convergence on such a timescale can be observed, and this is certainly not a matter of
the statistical uncertainty of conducting a single simulation.
The tendency of the mean first passage time to diverge instead of reaching a plateau
can also be observed for data published by Wedekind \cite[Figure 4.11 (bottom)]{Wedekind06}.
\end{multicols}

\begin{table}[h] \centering
\begin{tabular}{|rrr|rr|rrrr|}
\hline\hline
   $\superrho$ & $\superp$
 & $\supermu$
 & $\csizea$ & $\nuclrate(\csizea)$ [m$^{-3}$s$^{-1}$]
 & $\critsize(\textnormal{CNT})$
 & $\nuclrate(\textnormal{CNT})$ [m$^{-3}$s$^{-1}$]
 & $\critsize(\textnormal{LFK})$
 & $\nuclrate(\textnormal{LFK})$ [m$^{-3}$s$^{-1}$] \\
\hline

   $1.079$ & $1.077$ & $1.047$
 & $25$ & $3.3 \times 10^{33}$ & $629$
 & $2.3 \times 10^{29}$ & $1390$ & $2.7 \times 10^{15}$ \\
   $1.079$ & $1.078$ & $1.047$
 & $50$ & $6.5 \times 10^{32}$ & $607$
 & $3.1 \times 10^{29}$ & $1350$ & $5.7 \times 10^{15}$ \\

   $1.177$ & $1.112$ & $1.067$
 & $50$ & $4.8 \times 10^{32}$ & $222$
 & $1.5 \times 10^{32}$ & $596$ & $2.4 \times 10^{23}$ \\
   $1.177$ & $1.111$ & $1.066$
 & $75$ & $1.1 \times 10^{32}$ & $228$
 & $1.4 \times 10^{32}$ & $608$ & $1.7 \times 10^{23}$ \\

   $1.238$ & $1.133$ & $1.078$
 & $50$ & $2.4 \times 10^{33}$ & $140$
 & $7.7 \times 10^{32}$ & $416$ & $6.2 \times 10^{25}$ \\
   $1.238$ & $1.135$ & $1.079$
 & $75$ & $1.4 \times 10^{33}$ & $135$
 & $8.3 \times 10^{32}$ & $403$ & $1.0 \times 10^{26}$ \\
\hline\hline
\end{tabular}
\caption{Simulation results and theoretical values of nucleation rates for supersaturated
carbon dioxide at 285 K with \psat = 4712 kPa, \ndsat = 3.270 mol\slash{}l, and \planartension = 24.5 g/s$^2$;
for all values the threshold $\iota$ is lower than $\critsize$ according to theory}
\label{tabCO2T285K}
\end{table}

\begin{table}[h!] \centering
\begin{tabular}{|r||rrr|rr|rrrr|}
\hline\hline
     & $\temperature$ [K] & $\numdensity \,\left[\frac{\textnormal{mol}}{\textnormal{l}}\right]$
 & $\pressure$ [kPa]
 & $\csizea$ & $\nuclrate(\csizea) \left[\frac{1}{\textnormal{m}^3\textnormal{s}}\right]$
 & $\critsize(\textnormal{CNT})$
 & $\nuclrate(\textnormal{CNT}) \left[\frac{1}{\textnormal{m}^3\textnormal{s}}\right]$
 & $\critsize(\textnormal{LFK})$
 & $\nuclrate(\textnormal{LFK}) \left[\frac{1}{\textnormal{m}^3\textnormal{s}}\right]$ \\
\hline

\textbf{CH$_4$} & $106.0$ & $0.758$ & $503$
 & $25$ & $\mathbf{1.8 \times 10^{32}}$ & $22$
 & $\mathbf{1.6 \times 10^{29}}$ & $16$ & $\mathbf{9.9 \times 10^{31}}$ \\

 & $114.0$ & $0.851$ & $616$
 & $75$ & $\mathbf{2.7 \times 10^{31}}$ & $23$
 & $\mathbf{2.1 \times 10^{30}}$ & $ 19$ & $\mathbf{5.7 \times 10^{31}}$ \\
 & $114.0$ & $0.851$ & $614$
 & $150$ & $\mathbf{2.8 \times 10^{31}}$ & $23$
 & $\mathbf{2.0 \times 10^{30}}$ & $19$ & $\mathbf{5.4 \times 10^{31}}$ \\ 

 & $114.0$ & $0.925$ & $641$
 & $75$ & $\mathbf{5.8 \times 10^{31}}$ & $22$
 & $\mathbf{4.5 \times 10^{30}}$ & $18$ & $\mathbf{1.2 \times 10^{32}}$ \\
 & $114.0$ & $0.925$ & $629$
 & $150$ & $\mathbf{5.5 \times 10^{31}}$ & $23$
 & $\mathbf{3.2 \times 10^{30}}$ & $18$ & $\mathbf{8.5 \times 10^{31}}$ \\

 & $130.0$ & $1.432$ & $1022$
 & $700$ & $\mathbf{2.1 \times 10^{31}}$ & $31$
 & $\mathbf{3.2 \times 10^{31}}$ & $31$ & $\mathbf{3.1 \times 10^{31}}$ \\

 & $130.0$ & $1.606$ & $1095$ 
 & $75$ & $\mathbf{6.2 \times 10^{32}}$ & $26$
 & $\mathbf{1.3 \times 10^{32}}$ & $26$ & $\mathbf{1.7 \times 10^{32}}$ \\

 & $130.0$ & $1.693$ & $1148$
 & $25$ & $\mathbf{2.5 \times 10^{33}}$ & $24$
 & $\mathbf{3.2 \times 10^{32}}$ & $22$ & $\mathbf{6.2 \times 10^{32}}$ \\

 & $130.0$ & $1.780$ & $1167$
 & $25$ & $\mathbf{4.0 \times 10^{33}}$ & $23$
 & $\mathbf{4.1 \times 10^{32}}$ & $21$ & $\mathbf{7.8 \times 10^{32}}$ \\

\hline

\textbf{C$_2$H$_6$} & $176.5$ & $0.385$ & $455$
 & $25$ & $\mathbf{1.3 \times 10^{31}}$ & $16$
 & $\mathbf{4.5 \times 10^{30}}$ & $15$ & $\mathbf{2.7 \times 10^{31}}$ \\

 & $176.5$ & $0.400$ & $467$
 & $25$ & $\mathbf{2.0 \times 10^{31}}$ & $16$
 & $\mathbf{6.4 \times 10^{30}}$ & $14$ & $\mathbf{3.9 \times 10^{31}}$ \\

 & $280.0$ & $2.470$ & $3283$
 & $50$ & $3.9 \times \,10^{32}\,$ & $131$
 & $4.6 \times \,10^{32}\,$ & $322$ & $2.8 \times \,10^{26}\,$ \\
 & $280.0$ & $2.470$ & $3278$
 & $75$ & $1.7 \times \,10^{32}\,$ & $135$
 & $4.2 \times \,10^{32}\,$ & $329$ & $2.0 \times \,10^{26}\,$ \\

 & $280.0$ & $2.550$ & $3307$
 & $100$ & $1.7 \times \,10^{32}\,$ & $116$
 & $7.7 \times \,10^{32}\,$ & $290$ & $1.2 \times \,10^{27}\,$ \\

 & $280.0$ & $2.800$ & $3397$ 
 & $100$ & $\mathbf{1.8 \times 10^{33}}$ & $78$
 & $\mathbf{2.5 \times 10^{33}}$ & $207$ & $9.0 \times \,10^{28}\,$ \\

 & $280.0$ & $2.950$ & $3430$ 
 & $75$ & $\mathbf{3.3 \times 10^{33}}$ & $68$
 & $\mathbf{3.6 \times 10^{33}}$ & $186$ & $3.1 \times \,10^{29}\,$ \\
 & $280.0$ & $2.950$ & $3427$
 & $100$ & $\mathbf{1.9 \times 10^{33}}$ & $69$
 & $\mathbf{3.5 \times 10^{33}}$ & $188$ & $2.8 \times \,10^{29}\,$ \\

\hline

\textbf{CO$_2$} & $237.0$ & $1.700$ & $2283$
 & $75$ & $\mathbf{1.1 \times 10^{31}}$ & $50$
 & $\mathbf{9.4 \times 10^{30}}$ & $57$ & $\mathbf{7.7 \times 10^{29}}$ \\

 & $237.0$ & $1.750$ & $2317$
 & $75$ & $\mathbf{1.8 \times 10^{31}}$ & $48$
 & $\mathbf{1.6 \times 10^{31}}$ & $55$ & $\mathbf{1.3 \times 10^{30}}$ \\

 & $237.0$ & $1.850$ & $2322$
 & $300$ & $\mathbf{9.6 \times 10^{31}}$ & $47$
 & $\mathbf{1.7 \times 10^{31}}$ & $54$ & $\mathbf{1.5 \times 10^{30}}$ \\

 & $237.0$ & $2.000$ & $2333$
 & $300$ & $\mathbf{2.3 \times 10^{32}}$ & $46$
 & $\mathbf{2.0 \times 10^{31}}$ & $53$ & $\mathbf{2.0 \times 10^{30}}$ \\

 & $237.0$ & $2.450$ & $2499$
 & $75$ & $\mathbf{4.7 \times 10^{33}}$ & $37$
 & $\mathbf{1.4 \times 10^{32}}$ & $40$ & $\mathbf{4.2 \times 10^{31}}$ \\

 & $269.0$ & $3.120$ & $4142$
 & $25$ & $4.7 \times \,10^{33}\,$ & $81$
 & $4.7 \times \,10^{32}\,$ & $147$ & $6.7 \times \,10^{28}\,$ \\
 & $269.0$ & $3.120$ & $4131$
 & $75$ & $3.6 \times \,10^{32}\,$ & $83$
 & $4.2 \times \,10^{32}\,$ & $150$ & $5.1 \times \,10^{28}\,$ \\

 & $269.0$ & $3.800$ & $4350$
 & $50$ & $6.5 \times \,10^{33}\,$ & $54$
 & $2.6 \times \,10^{33}\,$ & $99$ & $6.7 \times \,10^{30}\,$ \\
 & $269.0$ & $3.800$ & $4343$
 & $75$ & $\mathbf{4.1 \times 10^{33}}$ & $55$
 & $\mathbf{2.5 \times 10^{33}}$ & $100$ & $5.8 \times \,10^{30}\,$ \\
\hline\hline
\end{tabular}
\caption{Simulation results and theoretical values of nucleation rates for supersaturated
methane, ethane, and carbon dioxide; bold values: threshold $\iota > \critsize$ according to theory}
\label{tabseveral}
\end{table}

\section{Comparison of theory and simulation}

\begin{figure}[h!]
\centerline{
   \includegraphics[width=5.67cm]{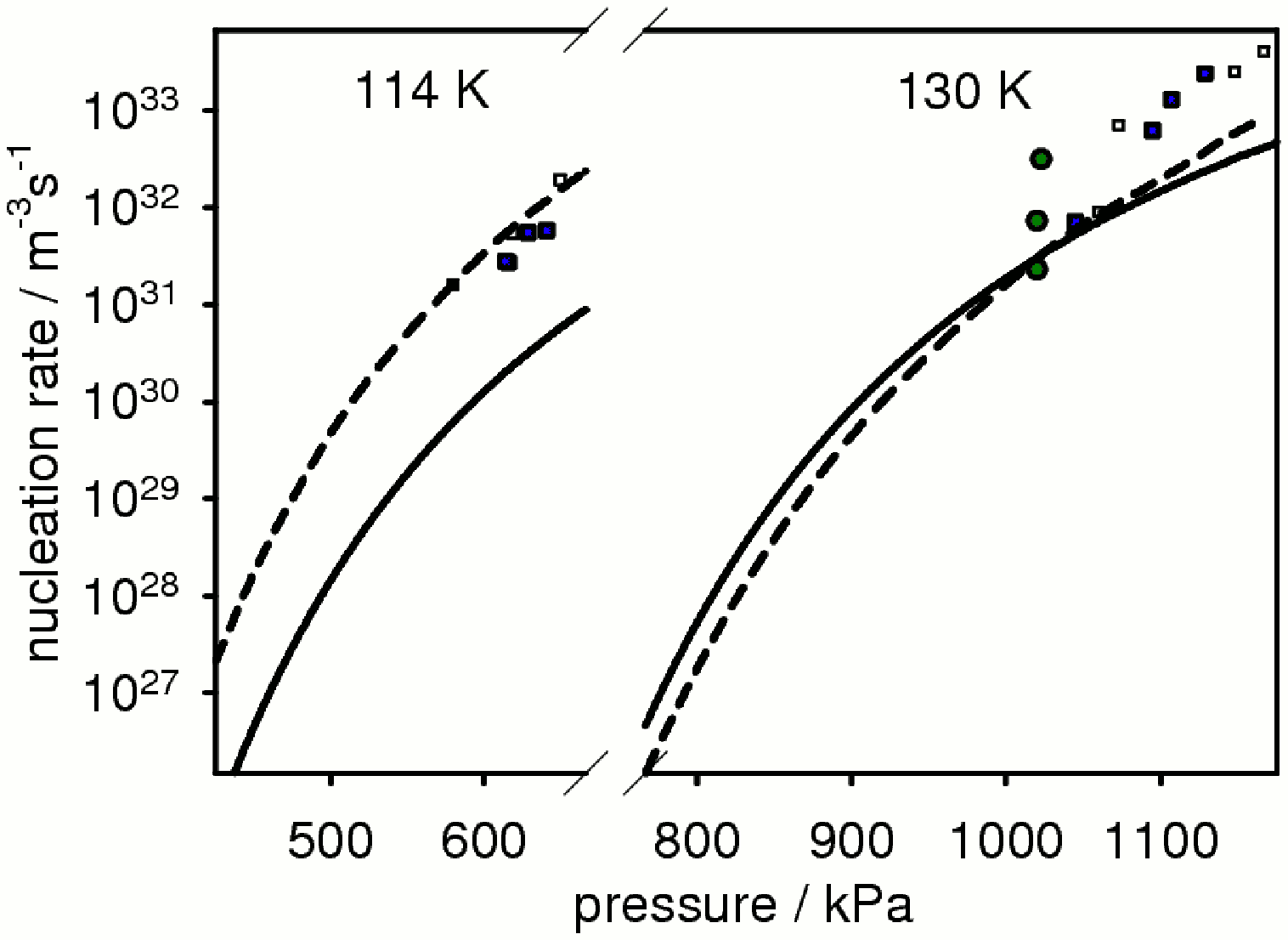}
      \quad\quad
   \includegraphics[width=5.67cm]{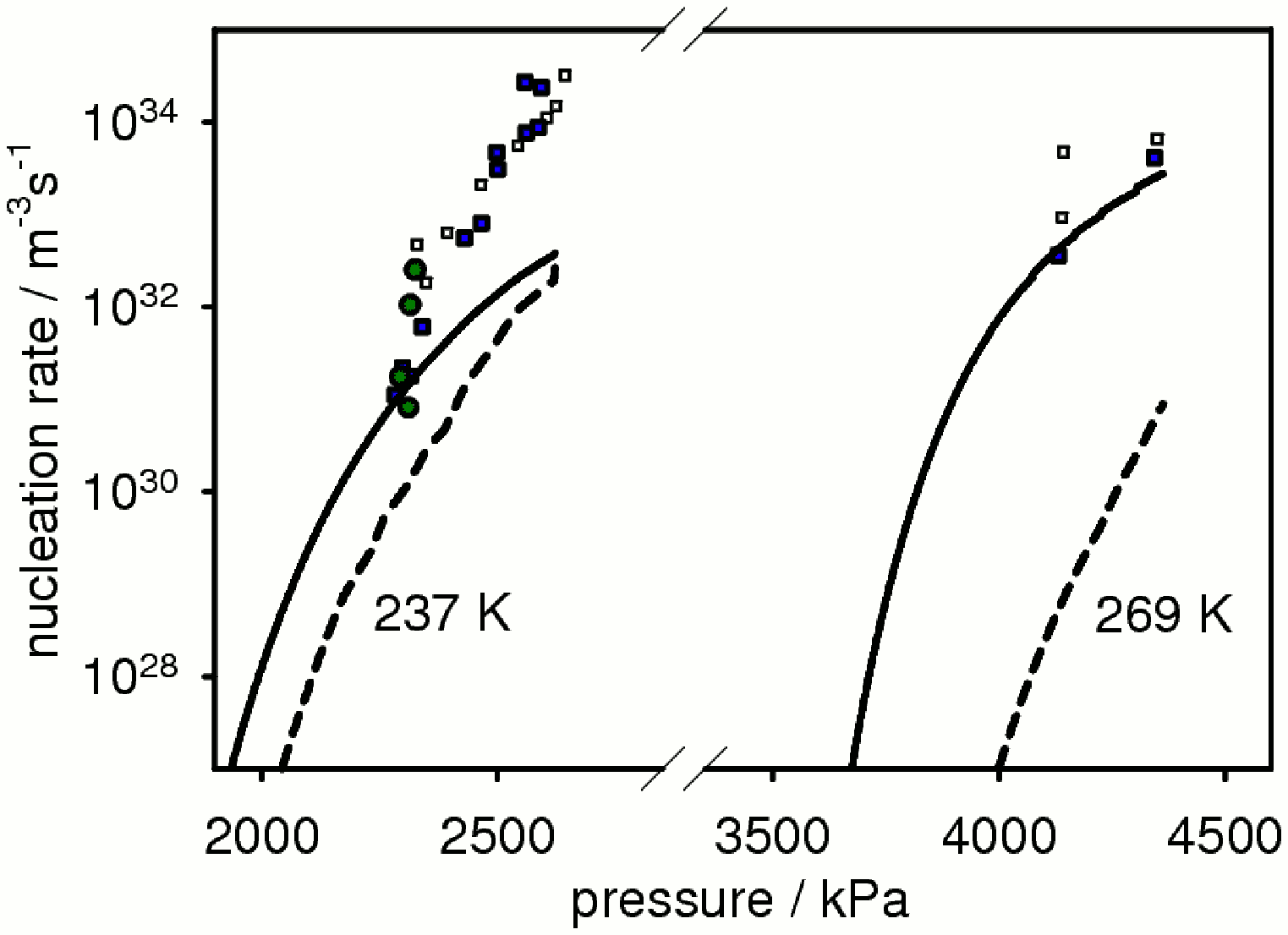}
      \quad\quad
   \includegraphics[width=5.67cm]{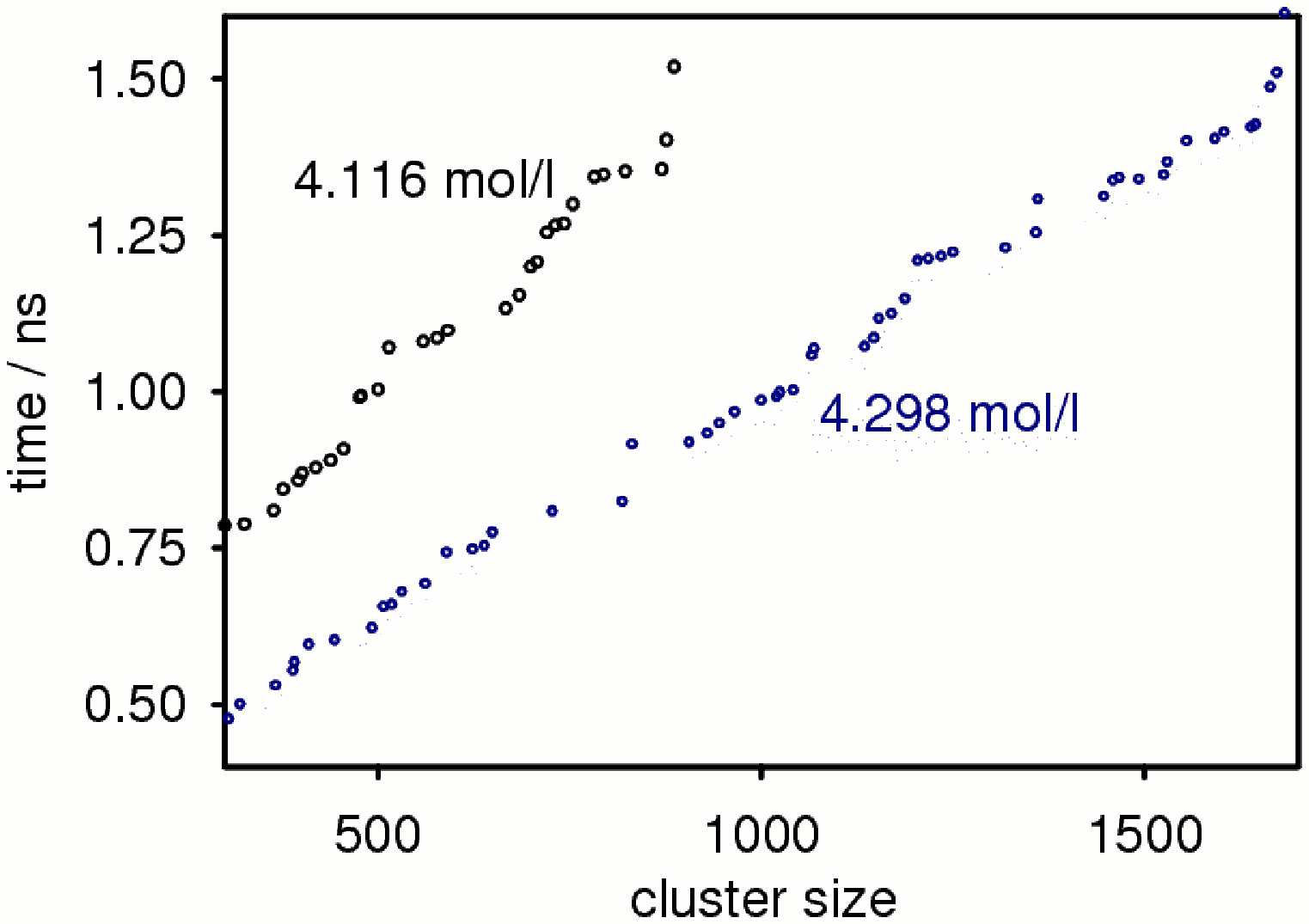}
}
\caption{
   \textbf{left} \,--\,
      Nucleation rate of CH$_4$ at 106, 114, and 130 K; small squares: \nuclrate(25),
      large squares: \nuclrate(75) and \nuclrate(150), circles: \nuclrate(225)
      and \nuclrate(700);
   \quad\textbf{center} \,--\,
      Nucleation rate of CO$_2$ at 237 and 269 K; small squares: \nuclrate(25) and \nuclrate(50),
      large squares: \nuclrate(75), circles: \nuclrate(300);
   \quad\textbf{right} \,--\,
      First passage time of clusters in CH$_4$ at 170 K with $\absnum = 250000$
}
\label{alephVIII}
\label{alephX}
\label{alephXII}
\end{figure}

\begin{multicols}{2}
The simulation results for $\nuclrate(\csizea)$ are compared to CNT and LFK
in Figures \ref{alephIX} (right) and \ref{alephVIII} (left and center)
as well as Tables \ref{tabCH4T170K}, \ref{tabCO2T253K}, \ref{tabCO2T285K}, and \ref{tabseveral}.

The correlation between $\pressure,\, \numdensity$ and
$\Delta\chempot$ at constant $\temperature$, which is necessary to evaluate the considered models,
was obtained from simulations of small supersaturated systems analogous to
those described in previous work \cite{LCVH05}. The dependence of $\pressure$ on
$\numdensity$ between data points was approximated by a linear fit. The resulting isotherms
were used to estimate the density $\numdensity(\pressure, \temperature)$ of the vapor,
which decreases over simulation time, to reflect that with a decreasing supersaturation,
nuclei should be expected to emerge at a lower rate -- note that the values of $\numdensity$ shown in the tables
correspond to the density of the entire system, not to the remaining vapor.
The isotherms were also applied to determine the second virial coefficient for the LFK model according
to Equation (\ref{eqn:virialLFK}), and the chemical potential
difference between the saturated and the supersaturated vapor according to
Equation (\ref{eqn:deltamu}). In Tables \ref{tabCH4T170K}, \ref{tabCO2T253K}, and \ref{tabCO2T285K}
the supersaturation with respect to the density, $\superrho = \numdensity\slash\satnumdensity(\temperature)$,
as well as the pressure are shown together with
\begin{equation}
   \supermu \,\,=\,\, \exp\left(\frac{\chempot(\pressure, \temperature)
      - \chempotsat(\temperature)}{\kboltz\temperature}\right),
\end{equation}
i.e.\ the supersaturation with respect to the chemical potential, where
$\satnumdensity(\temperature)$ and $\chempotsat(\temperature)$
refer to the saturated vapor at the given temperature. Occasionally, the identity
$\superrho = \superp = \supermu$ is assumed in the literature \cite{Ford04, HK03, Wedekind06};
in particular, it is used for the derivation of Equation (\ref{eqn:nucleationtheorem}), where $\superp$
replaces the more accurate $\supermu$.
However, near the spinodal this is always a bad approximation, since
$\partial\pressure\slash\partial\numdensity \to 0$ holds there by definition.

According to CNT, the size of the critical
nucleus for ethane at 280 K, 2.80 mol/l, and 3397 kPa is $\critsize = 78$,
cf.\ Table \ref{tabseveral}.
%
%
%
%
The overheating of the critical nucleus as predicted by CNT from Equation (\ref{eqnDeltaTcrit})
in this case, given that $\enthalpy = 8.3$ kJ/mol and $\zeldovich = 0.0089$,
amounts to $\DeltaTcrit = 1.4$ K; this corresponds to a nucleus temperature of 281.4 K
that is actually observed for large nuclei in the simulation, but not for $\csizea \approx 78$,
cf.\ Figure \ref{alephVII} (center).
From LFK we obtain $\critsize = 207$ as well as $\zeldovich = 0.0062$
and $\DeltaTcrit = 0.98$ K, which agrees well with the overheating observed for nuclei
containing $\csizea \approx 207$ molecules after a delay of 700 ps.

Values of $\nuclrate(\csizea)$ determined with the method of Yasuoka and Matsumoto are only
significant for $\csizea > \critsize$. Since the size of the critical nucleus can not be obtained by means
of this method, the theories are checked against their own predictions of $\critsize$: if the theoretical value of
$\critsize$ is smaller than the threshold used to evaluate the MD simulation, then simulation and theory
should be expected to agree. Such data are directly comparable -- they correspond to the highlighted
values in Tables \ref{tabCH4T170K}, \ref{tabCO2T253K}, \ref{tabCO2T285K}, and \ref{tabseveral}.

The values collected for the quadrupolar fluids show an eccellent agreement with CNT: all directly comparable
nucleation rates agree within one order of magnitude.
For methane, CNT significantly underestimates the nucleation rate at 106 K
and overestimates it at 170 K.
The predictions of $\nuclrate$ based on the LFK model are generally too low for carbon dioxide, with an error
between one and two orders of magnitude in the directly comparable cases. However, LFK predicts the
nucleation rate accurately for methane at 106, 114, and 130 K as well as for ethane at $176.5$ K.

Both theories are observed to deviate by about three orders of magnitude from certain directly
comparable $\nuclrate(\csizea)$ values: for methane at 106 K and 503 kPa, the method of Yasuoka and
Matsumoto yields $\nuclrate(25) = 1.8 \times 10^{32} \slash (\textnormal{m}^3\textnormal{s})$, while
CNT predicts $\nuclrate = 1.6 \times 10^{29} \slash (\textnormal{m}^3\textnormal{s})$, cf.\ Table
\ref{tabseveral}. For methane at 170 K and $\superp = 1.247$, cf.\ Table \ref{tabCH4T170K}, the 
nucleation rate according to the LFK model is
$\nuclrate = 6.7 \times 10^{29} \slash (\textnormal{m}^3\textnormal{s})$ as opposed
to $\nuclrate(225) = 2.7 \times 10^{32} \slash (\textnormal{m}^3\textnormal{s})$, obtained from an MD simulation.

\medskip

\noindent \textbf{Conclusion.} 
Molecular dynamics simulations of large nucleating systems
were conducted in the canonical ensemble and analyzed according to
the method of Yasuoka and Matsumoto.
It was shown that the average nucleus temperature, for a given nucleus size,
decreases over time, which leads to a considerable increase of the net growth rates during simulation.
Nucleation rate isotherms were obtained and compared to theoretical
predictions for methane at 106, 114, 130, and 170 K, for ethane at 176.5 and 280 K,
as well as for carbon dioxide at 237, 253, 269, and 285 K.
Nucleation rates from both CNT and LFK were observed to agree within three orders
of magnitude with those simulation results that can directly be compared to the theories.
In particular, CNT shows very small deviations for ethane and carbon dioxide over the entire
temperature range.
The LFK model consistently and significantly underpredicts nucleation rates at high temperatures, but agrees
very well for methane and ethane at low temperatures.
\end{multicols}

\noindent \textbf{Acknowledgment.} 
The authors thank Ralf Kible, Nicolas Schmidt, and
Jonathan Walter for fruitful discussions, Deutsche Forschungsgemeinschaft for funding
Sonderforschungsbereich 716 \qq{Dynamic Simulation of Systems with Large Numbers of Particles},
Landesstiftung Baden-W{\"u}rt\-temberg for funding project 688
\qq{Massiv parallele molekulare Simulation und Visualisierung der Keimbildung in Mischungen
f\"ur skalen\"ubergreifende Modelle,} as well as
the Simulation of Large Systems group at the Institute of Parallel and Distributed Systems
and the Numerics for Supercomputers group at the Institute of Applied Analysis and
Numerical Simulation, Universit\"at Stuttgart, for providing access to the \textit{mozart}
cluster. Some of the simulations were performed on the HP XC6000 super computer at the 
Steinbuch Centre for Computing, Karlsruhe, under the grant MMSTP and on the \textit{cacau}
cluster at the High Performance Computing Center Stuttgart (HLRS) under the grant MMHBF.

\begin{multicols}{2}
{\small
\bibliography{nuk2007jcp}
\bibliographystyle{ieeetr}
}
\end{multicols}

\end{document}